\documentclass[12pt]{article}

\usepackage{epsfig,amsmath}

\newcommand{\mupi}{\mu_\pi^2}
\newcommand{\mug}{\mu_G^2}
\newcommand{\rd}{\rho_D^3}
\newcommand{\rls}{\rho_{LS}^3}
\newcommand{\as}{\alpha_s}

\newcommand{\qqh}{\hat{q}^2}
\newcommand{\qzh}{\hat{q}_0}
\newcommand{\qq}{q^2}
\newcommand{\qz}{q_0}
\newcommand{\el}{E_\ell}

\newcommand{\GeV}{\,\mbox{GeV}}
\newcommand{\muwa}{\mu_{\scriptscriptstyle\rm WA}}

\newcommand{\lamh}{\hat{\Lambda}}
\def \be{\begin{equation}}
\def \ee{\end{equation}}
\newcommand{\bea}{\begin{eqnarray}}
\newcommand{\eea}{\end{eqnarray}}

\begin{document}
\begin{titlepage}

\begin{flushright}
DFTT-14/2007\\
UND-HEP-07-BIG\,06\\
\end{flushright}
\vskip 2cm

\centerline{\huge\bf\boldmath Inclusive semileptonic $B$ decays}
\vskip 2mm
\centerline{\huge\bf\boldmath  and the  
determination of $|V_{ub}|$}
\vskip 2cm

\begin{center}
{\bf 
  P.~Gambino$^a$, P.~Giordano$^a$, G.~Ossola$^b$,  and N.~Uraltsev$^{c,d}$\\[2mm]
$^a$\it   Dip.\ Fisica Teorica, Univ.\ di Torino, \& INFN  Torino,
I-10125 Torino, Italy\\
$^b$ Inst.\ of Nuclear Physics, NCSR "DEMOKRITOS", 15310 Athens, Greece.\\
$^c$   St.\ Petersburg Nuclear Physics Inst., Gatchina, St. Petersburg 188300, Russia\\
$^d$ Department of Physics, Univ.\ of Notre Dame, Notre Dame, IN 46556, USA}
\end{center}

\vskip 2cm

\begin{abstract}
We study the triple differential distribution of $B\to X_u \ell \nu$,
consistently including all perturbative and non-perturbative effects
through $O(\as^2\beta_0)$ and $O(1/m_b^3)$. The Fermi motion is
parameterized in terms of a single light-cone function for each
structure function and for any value of $q^2$, accounting for all
subleading effects. We discuss the problems and uncertainties related
to the high-$q^2$ tail and to Weak Annihilation effects. We work in
the {\it kinetic} scheme, a framework characterized by a Wilsonian
treatment with a hard cutoff $\mu\!\sim \! 1\:$GeV.  Our method is
illustrated with the extraction of $|V_{ub}|$ from some of the latest
experimental data, providing a detailed estimate of the theoretical
uncertainty.

\end{abstract}

\end{titlepage}


\section{Introduction}
The determination of the element $|V_{ub}|$ of the 
CKM matrix in semileptonic $B$ decays plays a central role 
in the search for flavour and CP violation beyond the Standard Model (SM).
Currently, global fits to flavour violating observables 
predict \cite{UTfit}
$$
|V_{ub}|_{UTfit}= (3.44\pm0.16)\times 10^{-3}
$$  
following the unitarity of the CKM matrix and assuming the validity of the SM. 
On the other hand, a direct determination of $|V_{ub}|$ is possible
based on the analysis of $b\to u \ell \nu$ decays, either in the inclusive
$B\to X_u \ell \nu$ or exclusive $B\to \pi \ell \nu$ channels.
Exclusive determinations rely on lattice QCD or light-cone sum rules
for the corresponding transition form factors and have improved in the 
last few years. Two  recent analyses give very consistent results:
$$
|V_{ub}|_{\rm excl}= (3.47 \pm 0.29 \pm 0.03)
\times 10^{-3}\, \cite{Flynn},\ \ \ \
|V_{ub}|_{\rm excl}=(3.5 \pm 0.4\pm 0.1) \times 10^{-3}\,{\cite{Ball}}.
$$

Inclusive decays based on a measurement of the total $b\to u \,\ell\nu$
decay rate potentially offer  the most accurate way to determine
$|V_{ub}|$. They are described by a local Operator Product Expansion
(OPE) in inverse powers of the $b$ quark mass \cite{1mb2}.  The OPE
has proved quite successful in the analysis of the moments of
various distributions in $B\to X_c \ell\nu $, leading recently to the
precise measurement of its dominant non-perturbative parameters, 
namely the $b$ and $c$ masses and the matrix elements of the
relevant dimension 5 and 6 local operators,  and to a 2\%
determination of $|V_{cb}|$ \cite{fits,BF}.  

In the case of charmless semileptonic decays experiments apply a
series of cuts to isolate the charmless decays that tend to destroy
the convergence of the local OPE. They introduce sensitivity to the
effects of Fermi motion of the heavy quark inside the $B$ meson, which
are not suppressed by powers of $1/m_b$ in the restricted kinematic
regions.  The Fermi motion is described in the OPE by a nonlocal
distribution function, whose lowest integer moments are given by the
expectation values of the same local operators we have encountered
earlier.

Fermi motion is of primary importance in another inclusive $B$ decay,
namely the radiative decay $b\to s \gamma$.  A dedicated OPE approach
accounting for the relation to the nonperturbative $B$-meson
parameters extracted from $B\to X_c\ell\nu$ was developed and applied
to the description of the photon energy moments with cuts in
Ref.~\cite{benson}. It proved quite successful in describing the
available $B\to X_s +\gamma$ data. The results of \cite{benson}
underline the importance of including subleading effects, going beyond
the leading-twist description of Fermi-motion. Another advantage of
the approach proposed in \cite{benson} is that it implements the
Wilsonian version of the OPE with a `hard' cutoff that separates the
perturbative and non-perturbative effects \cite{kinetic} and reduces
the significance of perturbative corrections. In this approach,
sometimes referred to as the {\it kinetic scheme}, the
non-perturbative parameters are also well-defined and perturbatively
stable. The contributions of soft gluons are absorbed into the
definition of the heavy quark parameters and of the distribution
function.

In this paper we develop an analogous approach for the case of the
triple differential distribution in $B\to X_u\,\ell\nu$ decays.  With
respect to the radiative decays, there are however a number of
complications due to the different kinematics.  In semileptonic decays
the invariant mass of the leptonic system, $q^2$, can vary up to
$M_B^2$. However the local OPE becomes problematic at high $q^2$,
where the effects of four-quark operators related to Weak Annihilation
(WA) also show up.  An accurate description of the physical spectra
requires a careful inclusion of these effects.

For what concerns perturbative corrections, the complete $O(\alpha_s^2
\beta_0)$ corrections to the triple differential rate have recently
been published \cite{Gambino:2006wk}.  To take advantage in our
framework of this new results and of the well-known $O(\as)$
corrections, we perform a new calculation of the real emission
contributions with the Wilsonian cutoff both at $O(\as)$ and
$O(\alpha_s^2 \beta_0)$. In this way, we include all perturbative
corrections to the triple differential rate through $O(\alpha_s^2
\beta_0)$ in the kinetic scheme: to the best of our knowledge, ours is
the most complete implementation of perturbative effects. 
The contribution of the $O(\alpha_s^2 \beta_0)$ effects turns out to be
numerically significant.  In our
treatment of perturbative corrections we do not resum Sudakov logs.
Indeed, in the framework with the hard cutoff, soft divergences are
absent by construction. The spectra still diverge at threshold due to
collinear divergences, but in a softer way.

Another element of our approach that is common with Ref.~\cite{benson}
concerns the distribution function, namely the inclusion of its power
corrections. We introduce the full finite-$m_b$ light-cone $b$-quark
distribution function.  Power effects enter it through the
power-suppressed terms in the OPE relations for its integer moments
and we take them into account through order $\Lambda_{\rm QCD}^3$
\cite{1mb3}.  At the level of power corrections the Fermi motion
effects cease to be universal: they depend on the process, and for
semileptonic decays they are function of $q^2$ and differ in the three
relevant structure functions $W_{1-3}(q_0,q^2)$.  We emphasize that we
do not split the distribution functions into leading and subleading
contributions.  Dealing with the full finite-$m_b$ distribution
functions we avoid calling upon a plethora of largely unconstrained
subleading functions. The latter typically are increasingly singular
in the end point, which is only an artifact of expanding in $1/m_b$
rather than a physical effect. We study in great detail the
dependence of the distribution functions on the assumed functional
form.

The significance of the effects of the Fermi motion proper in the
differential distributions fades away at larger $q^2$. However, at large
$q^2$ generic power corrections increase, and at some point even the
integrated moments cannot be described by the OPE: for $q^2$
approaching $m_b^2$ the decay process is no longer hard. This
signals the emergence of WA effects. In our approach the change of the
regime at certain $q^2$ automatically manifests itself, at least as
long as the $1/m_b^3$ effects are retained.  To avoid the
pathological behaviour of the OPE predictions in this kinematic
region, we model the high-$q^2$ tail in two different ways that 
preserve, for instance, the positivity of the differential rates.

Our approach is implemented in a numerical C++ code and we illustrate it
with the extraction of $|V_{ub}|$ from some of the latest experimental
results. Our results are compatible with 
the most recent Heavy Flavour Averaging Group (HFAG)
average for $|V_{ub}|$ from inclusive decays \cite{HFAG}
$$
|V_{ub}|_{\rm incl}=(4.34 \pm 0.16 \pm 0.25)\times 10^{-3}, \ \ \ \ 
|V_{ub}|_{\rm incl}= (4.31 \pm 0.17 \pm 0.35)\times 10^{-3},
$$ where the two values refer to two different theoretical frameworks
\cite{DGE,BLNP}, respectively, currently employed by HFAG. These
values are a few standard deviations away from the value preferred by
the global fit to the unitarity triangle and by the exclusive
determination. Our results,  however, indicate slightly 
larger uncertainties, especially due to WA effects.

The paper is organized as follows: in Section 2, after introducing
some notation, we discuss the perturbative
corrections to the triple differential semileptonic width in the
kinetic scheme with a hard cutoff.  In Section 3 we introduce the
distribution functions and their convolution with the perturbative
spectrum. In Section 4 we discuss different functional forms for the
light-cone functions.  Section 5 describes the problems encountered by
the OPE in the high-$q^2$ tail and presents two possible ways to
handle that kinematic region. Our results are discussed in Section 6,
where we extract $|V_{ub}|$ from recent experimental data using the
method developed in the previous sections. We also carefully discuss
the various sources of theoretical uncertainty. After the Conclusions,
the paper ends with three Appendices, containing certain details of our
calculations.

\section{Perturbative corrections with a  Wilsonian cutoff} 
\label{sec:wilsonian}
Our starting point is the triple differential distribution for $B\to
 X_u \ell\nu$ in terms of the leptonic variables $(\qz,\qq,\el)$ and
 of the three structure functions that are relevant in the case of
 massless lepton:
\begin{eqnarray} \label{eq:aquila_normalization}
\frac{d^3 \Gamma}{d\qq \,d\qz \,d\el}  &=&
\frac{G_F^2 |V_{ub}|^2}{8\pi^3}
 \Bigl\{ 
\qq W_1- \left[ 2\el^2-2\qz \el + \frac{\qq}{2} \right] W_2 \nonumber 
+  \qq (2\el-\qz) W_3 \Bigr\}\times \\&&
\quad\quad\quad\quad    \times \theta \left(\qz-\el-\frac{\qq}{4\el} \right) 
 \ \theta(\el) \ \theta(\qq) \ \theta(\qz-\sqrt{\qq}),
\end{eqnarray}
where $q_0$ and $E_\ell$ are the total leptonic and the charged lepton energies
in the $B$ meson rest frame and  $q^2$ is the leptonic invariant mass.
We will often use the normalized variables
\begin{equation}
\qzh = \frac{\qz}{m_b}, \qquad \qqh = \frac{q^2}{m_b^2},
\end{equation}
where $m_b$ is the $b$ quark mass. Within its range of validity, the
local OPE allows us to separate perturbative and power suppressed
non-perturbative contributions to the structure functions
\begin{equation}
\label{eq:pnp}
W_i(q_0,q^2,\mu) = m_b(\mu)^{n_i} \left[W_i^{pert}(\qzh,\qqh,\mu) + 
W_i^{pow}(\qzh,\qqh,\mu)\right] 
\end{equation}
with $n_{1,2}=-1$ and $n_3=-2$.  The quantities $ W_i^{pow}$ contain
the power corrections of the local OPE: their expressions through
${\cal O}(1/m_b^3)$ are quoted in the Appendix \ref{app:q0moments}.
In the context of the OPE with a Wilson cutoff $\mu$, the separation
between $W_i^{pert}$ and $ W_i^{pow}$ is controlled by $\mu$ and both
contributions are $\mu$-dependent.  The $\mu$-dependence, of course,
cancels out at each perturbative order in inclusive quantities like
the $q_0$-moments of $W_i(q_0,q^2,\mu)$. 

As already mentioned in the Introduction, we absorb the contributions
of soft gluons in the definition of heavy quark parameters and of the
distribution functions. Physical quantities are in principle
independent of the cutoff.  The presence of the cutoff introduces
several modifications in the perturbative structure functions. They
have been studied at $q^2=0$ for radiative decays in \cite{benson}.
In our case the structure functions take the form, through
$O(\alpha_s^2 \beta_0)$:
\begin{eqnarray} \label{Wpert}
W_i^{pert} (\qz,\qq,\mu) &=&
\left[ W_i^{\rm tree}(\hat{q}^2) + 
C_F \frac{\alpha_s(m_b)}{\pi} V_i^{(1)}(\hat{q}^2,\eta) + 
C_F \frac{\alpha^2_s \beta_0}{\pi^2} V_i^{(2)}(\hat{q}^2,\eta) \right]
\delta(1+\hat{q}^2-2\hat{q}_0) \nonumber \\
& & + \quad C_F \frac{\alpha_s(m_b)}{\pi} \left[R_i^{(1)}(\hat{q}_0,\hat{q}^2,\eta) 
+\frac{\alpha_s \beta_0}{\pi} R_i^{(2)}(\hat{q}_0,\hat{q}^2,\eta) 
\right] \theta(1+\hat{q}^2-2\hat{q}_0) \nonumber \\
& & + \quad C_F \frac{\alpha_s(m_b)}{\pi}\left[ B_{i}^{(1)}(\qqh,\eta)+
\frac{\alpha_s \beta_0}{\pi}  B_{i}^{(2)}(\qqh,\eta)\right] 
\delta'(1+\hat{q}^2-2\hat{q}_0),
 \label{cutSp}
\end{eqnarray}
where
\begin{equation}
\eta = \mu / m_b.
\end{equation}
We always assume $0 < \eta < 1/2$,
and $\beta_0=11-\frac23 n_f$ with $n_f$ the number of light
active flavours. In the numerics, we will set $n_f=3$. The derivative of the 
Dirac's delta in \eqref{cutSp} is taken wrt to its argument.

The normalization of the $W_i$ is such that
\begin{equation}
W_1^{\rm tree}(\hat{q}^2) = 1-\qqh, \qquad W_2^{\rm tree}(\hat{q}^2) = 4, \qquad W_3^{\rm tree}(\hat{q}^2) = 2.
\end{equation}
\begin{figure}[t]
\begin{center}
\mbox{\epsfig{file=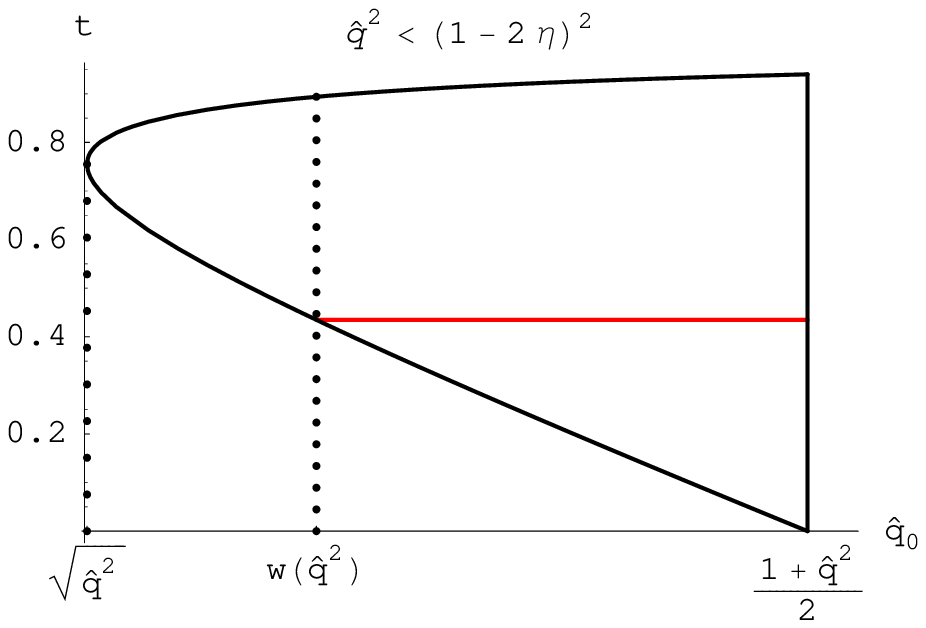,width=7.5cm}}
\mbox{\epsfig{file=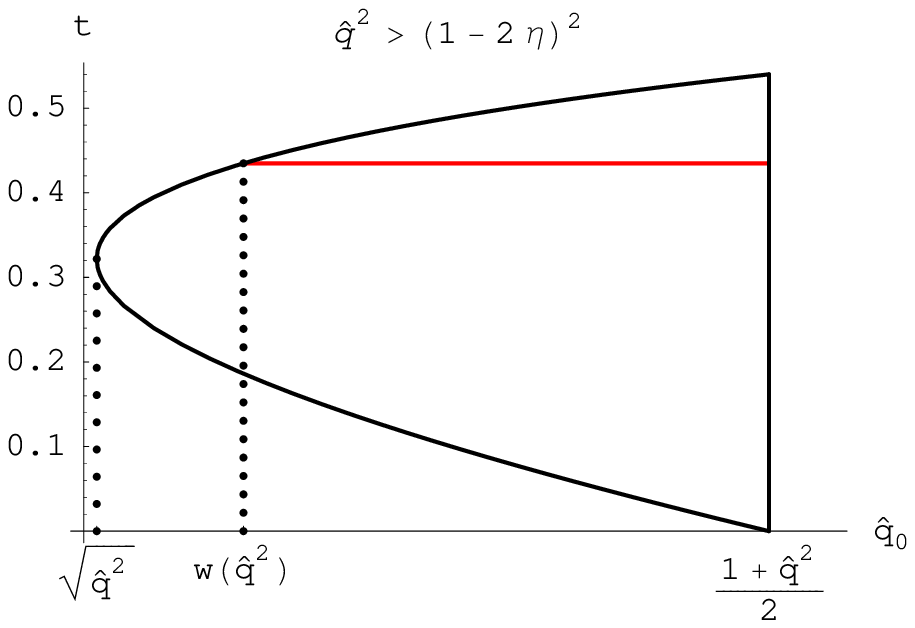,width=7.5cm}}
\end{center}
\caption{\sf The real gluon emission energy in the presence of a cutoff, with
$t=2 E_g/m_b$. In the on-shell decay amplitude $t$ is constrained
to lie within the black solid line. The red horizontal line represents the 
cutoff $t>2 \eta$. The left (right)  plot is at low (high) $q^2$. 
See the text.
}\label{gluonen}
\end{figure}
 
We have computed the real gluon emission contributions $
R_i^{(1,2)}(\hat{q}_0,\hat{q}^2,\eta)$ restricting the phase space
integration in $b\to u\, W^*\, g$ to gluons with energies larger than
the cutoff $\mu$.  The effect of the cutoff is to remove the infrared
divergence, softening the divergence of the form factors at the
endpoint, where collinear divergences are still present.  The
calculation of BLM corrections has been performed using the
 technique with a massive
gluon and integrating over the gluon mass, a standard procedure
described, for instance, in \cite{Aquila:2005hq}, that can be extended
to compute and resum $O(\as^n \beta_0^{n-1})$ corrections.  We have
also reproduced all results of the analogous calculation performed in
\cite{benson} for the case of $b\to s\gamma$.

The functional form of  the real contributions
$ R_i^{(k)}(\hat{q}_0,\hat{q}^2,\eta)$ depends on the 
region of the parameter space:
\begin{equation} \label{real}
R_i^{(k)}(\hat{q}_0,\hat{q}^2,\eta) = 
\widetilde{R}_i^{(k)}(\hat{q}_0,\hat{q}^2)
\, \theta \left(w-\qzh \right)\theta \left(1-2\eta-\sqrt{\qqh} \right)+ 
R_{i}^{cut,(k)}(\hat{q}_0,\hat{q}^2,\eta)
\, \theta \left( \qzh - w \right)
\end{equation}
where  
\be
w \equiv w(\qqh,\eta) = \frac12-\eta+\frac{\qqh}{2\,(1-2\eta)}
\ee
and $\widetilde{R}_i^{(1,2)}(\hat{q}_0,\hat{q}^2)$ are the real emission 
contributions in the on-shell scheme (without cutoff, $\mu=0$) 
calculated at $O(\alpha_s)$ \cite{dfn} and  
$O(\alpha_s^2\beta_0)$ \cite{Gambino:2006wk}.
Eq.(\ref{real}) can be understood from  Figs.~\ref{gluonen},
where the integration range for the gluon energy $E_g$ in the $b$ rest frame
is represented in terms of $t =2 E_g/m_b$ and $\qzh$ at fixed $\qqh$. 
The on-shell kinematics implies 
$1-\qzh-\sqrt{\qzh^2-\qqh}<t<1-\qzh+\sqrt{\qzh^2-\qqh}$.
At low $\qqh$ (left plot) the cutoff is irrelevant for
$\qzh<w(\qqh,\eta)$  while it modifies the spectrum for 
$w(\qqh,\eta)<\qzh<(1+\qqh)/2$. For large enough values of $\qqh$,
$\qqh>(1-2\eta)^2$,  the energy of the gluons  is always below the cutoff 
unless $w(\qqh,\eta)<\qzh$ and the spectrum is affected at all $\qzh$ 
(right plot). At even higher lepton invariant mass, $\qqh>(1-2\eta)$,
 gluon emission is completely inhibited. 
\begin{figure}[t]
\begin{center}
\mbox{\epsfig{file=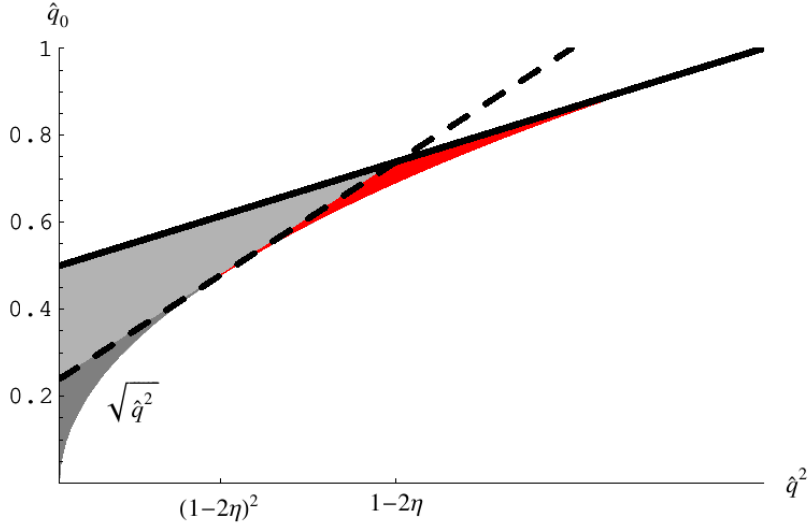,width=11cm}}
\end{center}
\caption{\sf 
Different domains for $R_i^{(k)}$ in the  $(\qzh,\qqh)$ plane. The solid 
and dashed straight lines represent $(1+\qqh)/2$ and $w(\qqh,\eta)$, 
respectively. 
Regions I, II, and III comprised between these two lines and 
the lower bound $\sqrt{\qqh}$
 are shown in dark gray, light gray and red (see text).
We have used   $\mu=1.2$~GeV,  $m_b=4.6$~GeV. 
}\label{q0-q2}
\end{figure}
One therefore 
identifies three regions in the $(\qzh,\qqh)$ plane that are displayed 
with different colors in  Fig.~\ref{q0-q2}:
(I) where the cutoff does not modify real emission (dark gray), (II) where the 
cutoff modifies the real emission (light gray), (III) where the 
presence of the cutoff inhibits completely real gluon emission
(red).

\begin{figure}[t]
\begin{center}
\mbox{\epsfig{file=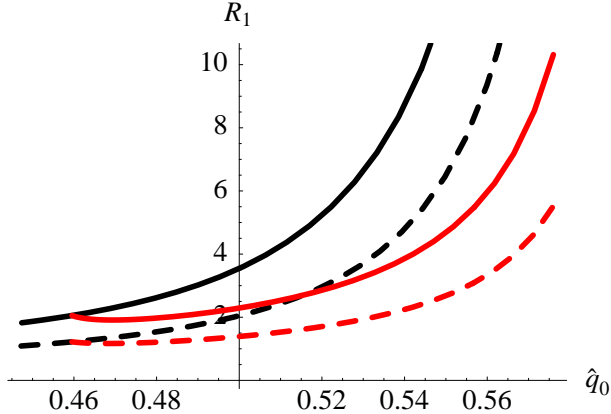,width=9cm}}
\end{center}
\caption{\sf The real gluon contribution $R^{(1)}_1$ (dashed curves) and 
the sum $R^{(1)}_1+ \as\beta_0/\pi  \, R^{(2)}_1$ (solid curves)
as functions of $\qzh$ at $\qqh=0.2$  with (red) and without 
(black) a cutoff $\mu=1$~GeV on the energy of the gluons.
}\label{cutoff-spect}
\end{figure}

Explicit expressions for $R_i^{(1)}(\hat{q}_0,\hat{q}^2,\eta)$ are
reported in the Appendix A. We do not give those for $
R_i^{(2)}(\hat{q}_0,\hat{q}^2,\eta)$ which are relatively lengthy, but
they can be found in the computer code.  Fig.~\ref{cutoff-spect} shows
the $\qzh$ dependence of $W_1$ with and without the cutoff.  Here and
in all numerical examples of this section, we assume $\mu=1$~GeV,
$m_b(1~{\rm GeV})=4.6$~GeV.  As anticipated, the structure function is
unchanged for $\qzh<w$, and diverges less severely close to the
endpoint.  Therefore we do not perform any resummation, unlike
\cite{DGE,BLNP,ugo}.

A direct calculation of the virtual contributions $V_i^{(1,2)}$
in the presence of the cutoff $\mu$ is more cumbersome. However, since their
expressions in the absence of the cutoff (on-shell scheme) are known from
\cite{dfn,Gambino:2006wk}, one can infer
their expressions at arbitrary $\mu$ from  the requirement that
physical quantities be independent of the cutoff at each perturbative order
when both power and perturbative corrections are consistently included.
In particular, following the argument given in the previous section,
we require the $\mu$-independence of the integral over 
$q_0$ of each structure function:
\begin{equation} \label{eq:matching_0_mom}
\int_{-\infty}^{+\infty} \ dq_0 \,W_i(q_0,q^2,0) = 
\int_{-\infty}^{+\infty} \ dq_0 \,W_i(q_0,q^2,\mu) 
+ {\cal O}(\as^2).
\end{equation} 

In order to extract the $\mu$-dependence of $V_i$, we therefore
consider the various sources of cutoff dependence 
in Eq.~(\ref{eq:matching_0_mom}).
The renormalization of the non-perturbative parameters and of $m_b$ 
is known to induce  a cutoff-dependence of the form \cite{kinetic,imprecated}:
\begin{equation}\label{kindef}
m_b(0) = 
m_b(\mu) + \left[\bar\Lambda(\mu)\right]_{\rm pert} 
+  \frac{\left[\mu^2_\pi(\mu) \right]_{\rm pert}}
{2 m_b(\mu)},
\end{equation}
\begin{equation}
\mupi(0) = \mupi(\mu) - [\mupi(\mu)]_{\rm pert}\label{kindef2}
\end{equation}
\begin{equation}
\rd(0) = \rd(\mu) - [\rd(\mu)]_{\rm pert},\label{kindef3}
\end{equation}
with\footnote{These expressions actually refer to the asymptotic value
of $\mupi$, namely in the infinite $m_b$ limit.  In general we  
employ a definition of $\mupi$ and of the other OPE
parameters at finite $m_b$, but we actually choose 
to neglect $O(\mu^3)$ terms in $[\mupi(\mu)]_{\rm pert}$ and 
$ [\mug(\mu)]_{\rm pert}$ in order to be 
consistent with Ref.~\cite{btoc} and with the way the global fit
for the determination of these parameters is performed in
\cite{BF}. This amounts to an {\it ad-hoc} perturbative redefinition 
of $\mupi$ and $\mug$. }
\begin{eqnarray} 
\left [\bar\Lambda (\mu) \right ]_{\rm pert} &=&
\frac {4}{3}C_F \frac {\alpha_s(m_b)}{\pi} \mu \left[1+ 
\frac{\alpha_s\beta_0}{2\pi}\left(\ln \frac{m_b}{2\mu}+\frac83  \right)\right]
,\nonumber\\
\left [\mu_{\pi}^2 (\mu) \right ]_{\rm pert} &=&
C_F \frac{\alpha_s(m_b)}{\pi} \mu^2 \left[1+ 
\frac{\alpha_s\beta_0}{2\pi}\left(\ln \frac{m_b}{2\mu}
+\frac{13}6  \right)\right],\nonumber\\
\left[ \rd(\mu) \right]_{\rm pert} &=& \frac{2}{3} C_F \frac{\alpha_s(m_b)}{\pi} 
\mu^3\left[1+ 
\frac{\alpha_s\beta_0}{2\pi}\left(\ln \frac{m_b}{2\mu}+2  \right)\right].
\end{eqnarray}

It is worth recalling that the perturbative shifts in
Eqs.~(\ref{kindef}-\ref{kindef3}) are not only conceptually, but also
numerically quite important: using the current best experimental
determination $\mupi(1\,{\rm GeV})\approx 0.40\, {\rm GeV}^2$, for
instance, we see that the perturbative shift $[\mupi(1\,{\rm
GeV})]_{\rm pert}$ amounts to almost 40\% of that.  We now write the
perturbative contributions to the structure functions in the following
way:
\begin{eqnarray}
R_i^{(1,2)}(\qzh,\qqh,\eta) &=& \widetilde{R}_i^{(1,2)}(\qzh,\qqh)+\delta R_i^{(1,2)}(\qzh,\qqh,\eta),\\ 
V_i^{(1,2)}(\qqh,\eta) &=& 
\widetilde{V}_i^{(1,2)}(\qqh)+\delta V_i^{(1,2)}(\qqh,\eta),
\end{eqnarray}
where $\widetilde{V}_i^{(1,2)}(\qqh)$ are the soft-virtual 
contributions in the on-shell scheme (without cutoff) 
calculated at $O(\alpha_s)$ \cite{dfn} and  
$O(\alpha_s^2\beta_0)$ \cite{Gambino:2006wk} (see Appendix A).

\begin{figure}[t]
\begin{center}
\mbox{\epsfig{file=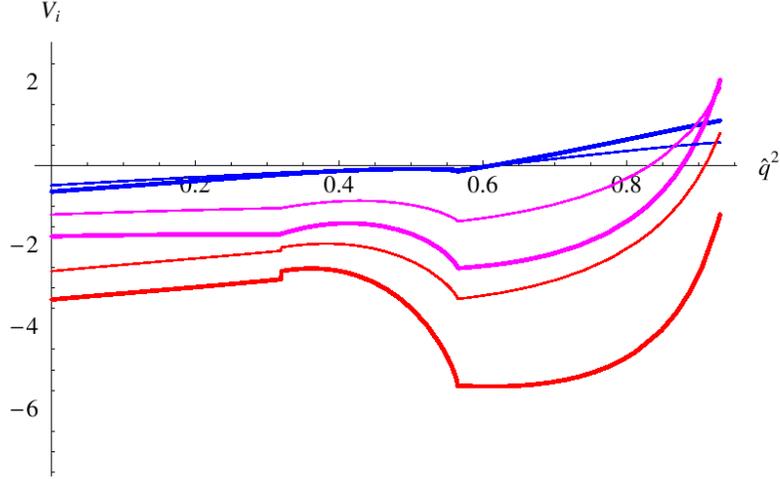,width=11cm}}
\end{center}
\caption{\sf Virtual contributions to the structure functions 
 at $O(\as)$, $V_i^{(1)}(\qqh,\mu)$ (thin lines) 
and through   $O(\as^2 \beta_0)$, $V_i^{(1)}+ \frac{\as\beta_0}{\pi}V_i^{(2)} $
 (thick lines) as functions of $\qqh$ for $\mu=1$~GeV and $\alpha_s(m_b)=0.22$.
The blue, red, and magenta lines correspond to 
$i=1,2,3$, respectively. }\label{Vip}
\end{figure}

The matching condition \eqref{eq:matching_0_mom} is satisfied if and only if
\begin{equation}\label{deltaVi}
\delta V_i^{(k)}(\qqh,\eta) = S_i^{(k)}(\qqh,\eta) - 2
\int_{\sqrt{\qqh}}^{\frac{1+\qqh}2} \ d\qzh \ \delta R_i^{(k)}(\qzh,\qqh,\eta) \ \theta(1+\qqh-2\qzh),
\end{equation}
where the range of the integral on the rhs is fixed by the decay kinematics.
It is best performed numerically, 
while the $S_i^{(k)}(\qqh,\eta)$ are  given by
\begin{align}
 \frac{C_F \as(m_b)}{\pi} &\left[S_i^{(1)}
+\frac{\as\beta_0}{\pi} S_i^{(2)}
\right] \frac{  m_b^{n_i+1}}{2} =  \left[ 
 \left[\bar\Lambda(\mu)\right]_{\rm pert} 
+ \frac{\left[\mu^2_\pi(\mu) \right]_{\rm pert}}
{2 m_b} \right] \frac{\partial M_i^{(0),{\rm tree}}\left(\frac{\qq}{ m_b^2}\right)}{\partial m_b} 
\nonumber\\
& \hspace{2.5cm} -  \left[ \left[\mu^2_\pi(\mu) \right]_{\rm pert} \frac{\partial}{\partial \mu_\pi^2}
+ \left[\rd(\mu) \right]_{\rm pert} \frac{\partial}{\partial \rd} \right] 
M_i^{(0),pow}\left( \qqh \right). 
\end{align}
The tree-level and power corrections contributions 
to the zeroth moments, $M_i^{(0),{\rm tree}}$ and $M_i^{(0),pow}$,
  can be found in Appendix B. 
The resulting factors $S_i^{(k)}$ read:
\begin{eqnarray}
&& S_1^{(1)} = \frac{8}{3}\qqh \eta 
+ \frac{\qqh-2}{3}\eta^2,
 \quad\quad
S_1^{(2)} = \frac{32 \qqh }{9}\eta+\frac{13(\qqh-2)}{36} \eta^2
-S_1^{(1)}\, \frac{\ln {2\eta}}2  , 
\nonumber\\
&& S_2^{(1)} = 0,\hspace{3.5cm} S_2^{(2)} = 0  ,
\\
&& S_3^{(1)} = -\frac{8}{3}\eta+\frac{2}{9}\eta^3,
\hspace{1.7cm}
S_3^{(2)} =-\frac{32 }{9 }\eta+\frac{2 }{9}\eta^3 
- S_3^{(1)}\,\frac{\ln {2\eta}}2 .
\nonumber
\end{eqnarray}
\begin{figure}[t]
\begin{center}
\mbox{\epsfig{file=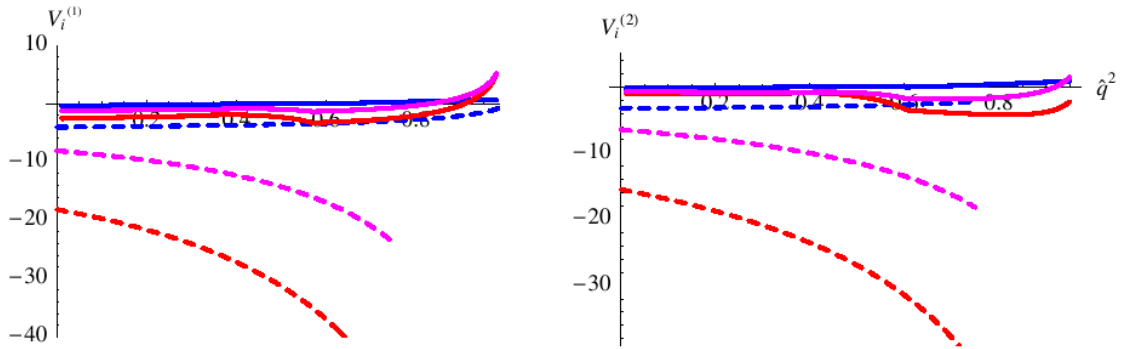,width=15cm}}
\end{center}
\caption{\sf Virtual contributions to 
$V_i^{(1)}(\qqh,\mu)$  (left) and $V_i^{(2)}(\qqh,\mu)$ (right) 
and their pole mass scheme counterparts $\widetilde{V}_i^{(1,2)}=
V_i^{(1,2)}(\qqh,0)$ (dashed lines) 
as functions of $\qqh$ for $\mu=1$~GeV and $\alpha_s(m_b)=0.22$.
The blue, red, and magenta lines correspond to 
$i=1,2,3$, respectively. \label{Vi}}
\end{figure}
The complete virtual contributions to the three structure functions
are shown in Fig.~\ref{Vi} for $\mu=1$~GeV, at $O(\alpha_s)$ and
$O(\alpha_s^2 \beta_0)$.  There is a strong suppression of the virtual
contributions with respect to the case without cutoff and the BLM
corrections are typically smaller in relative terms, as it can be
expected since the cutoff increases the typical gluon energy.  It is
worth noting that, following our discussion of
Figs.~\ref{gluonen}-\ref{q0-q2}, there are three regions of $\qqh$ in
the calculation of the integrals of $\delta R_i^{(1,2)}$ in
\eqref{deltaVi}.  This is illustrated in Fig.~\ref{Vip} for the case
of the virtual contributions $V_i$.  For $\qqh>(1-2\eta)^2$ terms
non-analytic in $\mu=0$ appear: the radius of convergence of the
$\mu$-expansion decreases for increasing $\qqh$ and one does not
expect the $\mu$-independence to hold when one includes only a few
higher dimensional operators, as done above.  This is actually related
to the poor convergence of the OPE at high $q^2$ that we will discuss
more in detail later on. In practice, our choice of a cutoff close to
1 GeV implies a rather small value $(1-2\eta)^2\approx 0.32$ for the
first threshold in $\qqh$.  A high cutoff might indeed magnify at
moderate $q^2$ the contribution of higher dimensional operators that
are inevitably important in the high-$q^2$ tail.  For this reason in
our implementation we keep the complete $\eta$-dependence of $\delta
V_i$, but one can examine different options to study the theoretical
uncertainty.

There is one point left to clarify in Eq.~\eqref{Wpert}, namely the
presence of a $\delta'$ term.  This is related to the difference that
occurs between the kinetic mass of the {\it b} quark
and the rest-energy that determines the kinematic end-point
in the decay of the heavy quark
\cite{benson}. This difference implies a finite perturbative shift of
the endpoint, which manifests itself in Eq.~\eqref{Wpert} as a
derivative of the Dirac delta.
The values of $B_{i}^{(1,2)}$ can be calculated explicitly, but  
again we choose to infer their expressions from the $\mu$-independence.  
To this end we require the cutoff-independence of the first $q_0$  moment 
(the only moment affected by the presence of a $\delta'$).
A procedure analogous to the one described for the virtual corrections 
yields:
\begin{align}\label{Biexpr}
 \frac{C_F \as}{\pi}&\left[B_{i}^{(1)}(\qqh,\eta)
+ \frac{\as\beta_0}{\pi} B_i^{(2)}(\qqh,\eta)\right]
 = \frac{2\,(1-\qqh)}{ m_b^{2+n_i}} \left[ 
\left[\bar\Lambda(\mu)\right]_{\rm pert} 
+ \frac{\left[\mu^2_\pi(\mu) \right]_{\rm pert}}  
{2 m_b} \right] M_i^{(0),{\rm tree}}(\qqh) \nonumber\\
& \quad\quad\quad - \frac4{ m_b^{2+n_i}}\left[
 \left[\mu^2_\pi(\mu) \right]_{\rm pert} \frac{\partial}{\partial \mu_\pi^2}
+ \left[\rd(\mu) \right]_{\rm pert} \frac{\partial}{\partial \rd} \right]
I_i^{(1),\text{pow}}(\qqh) \\
&\quad\quad\quad -  4 \, \frac{C_F \as}{\pi}
\int_{\sqrt{\qqh}}^{\frac{1+\qqh}2}
d\qzh \left(\qzh-\frac{1+\qqh}{2}\right) \left[
\delta R_i^{(1)}  +\frac{\as\beta_0}{\pi} \delta R_i^{(2)} 
\right],
\nonumber
\end{align}
where the power corrections to the first 
central moments of the structure functions, $I_i^{(1),\text{pow}}(\qqh)$,
are given in the Appendix B.
In the region $\qq<(1-2\eta)^2$, expanding  in $\eta$ up to $O(\eta^4)$, 
we find 
\begin{eqnarray}
B_{1}^{(1,2)}&=& \frac{1-\qqh}2 \, B_{3}^{(1,2)}
,\nonumber\\
B_{2}^{(1)}&=& \frac{4}{3}(1-\qqh) \ \eta+4 \ \eta^2+ \frac{4}{9}(1+5\qqh) \ \eta^3, \nonumber\\
B_{3}^{(1)} &=& \frac{2}{3}(1-\qqh) \ \eta+2 \ \eta^2+\frac{2}{9}(-7+5\qqh) \ \eta^3.\label{eqB1}
\end{eqnarray}
\begin{eqnarray}
B_{2}^{(2)}&=&
1.444 (1-\qqh) \eta+ (4.691-0.0544\qqh +0.092\hat{q}^4) \eta ^2\nonumber\\&&
-( 0.7517  -4.353\qqh +5.612\hat{q}^4 )\eta^3
 -B_2^{(1)}\, \frac{\ln {2\eta}}2  , \nonumber\\
B_{3}^{(2)} &=& 0.7219 (1 - \qqh) \eta + 
    (2.353 - 0.044\qqh + 0.05\hat{q}^4)\eta^2 \nonumber\\
&& - 
     (2.772 - 2.65  \qqh + 2.8\hat{q}^4) \eta^3  
-B_3^{(1)}\, \frac{\ln {2\eta}}2 \nonumber .
\end{eqnarray}
\begin{figure}[t]
\begin{center}
\mbox{\epsfig{file=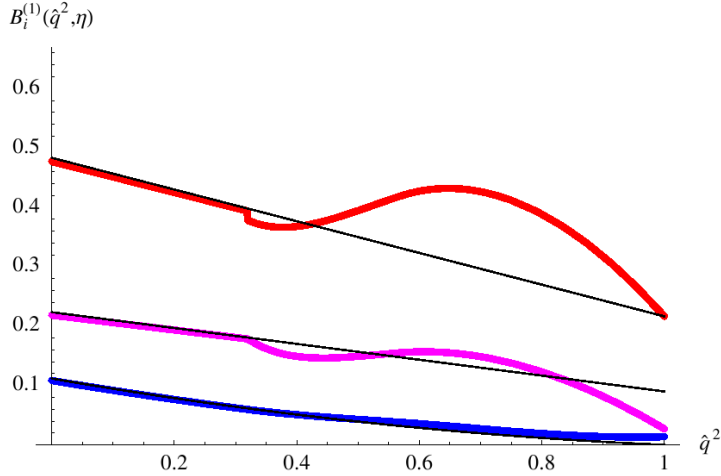,width=10cm}}
\end{center}
\caption{\sf 
The coefficients $B_{i}^{(1)}$
as functions of $\qqh$ for $\mu=1$~GeV. 
The blue, red, and magenta thick lines correspond to 
$i=1,2,3$, respectively. The thin black lines are the corresponding 
$\eta$ expansions given in Eq.~\eqref{eqB1}.}\label{B1}
\end{figure}
The $B_{i}^{(2)}$ are given as high precision 
approximate formulas that reproduce the numerical results within 1\%. 

As noted above, non-analyticity in $\eta$ is significant at high $\qqh$.
This is again illustrated in Fig.~\ref{B1} for the case of
$B_i^{(1)}$. The figure compares the full $\eta$-dependence with the
expansion in $\eta$ through $O(\eta^3)$.  For $\qqh>(1-2\eta)^2$ there
is no reason to expect $B_i$ to be well described by an expansion in
$\eta$.  It turns out that non-analyticity is a minor effect for
$i=1,3$ up to $\qqh\sim 0.8$, while for $i=2$ it becomes relevant
already at $\qqh\sim 0.5$. On the other hand,  $W_2$ is
kinematically suppressed in the triple differential rate at high
$q^2$.  In our practical implementation of perturbative corrections,
we employ only the expanded formulas \eqref{eqB1} for $B_i^{(k)}$: the
consequent mismatch at high $q^2$ manifests itself as a mild
$\mu$-dependence of the physical quantities that we compute and will
be taken into account in our estimate of theoretical uncertainties.

\section{Distribution functions in $B\to X_u l \nu$} 
\label{sec:SF_derivation}
The resummation of the leading twist effects in semileptonic $B$ decays close 
to threshold has been first studied long ago  \cite{Bigi:1993ex}. The
structure functions are expressed as the convolution of their tree-level 
expressions with the light-cone distribution function $F(k_+)$ whose support 
lies below $\bar\Lambda=M_B-m_b$, where $M_B$ is the $B$ meson mass:
\begin{equation} \label{eq:conv1}
W_i(q_0,q^2) = m_b^{n_i} \,\int dk_+ \ F(k_+) \ W_i^{(0)}
\left[ q_0 - \frac{k_+}{2} \left( 1 - \frac{q^2}{m_b^2} \right), q^2 \right]
\end{equation}
where 
$$ W_i^{(0)}(q_0,q^2)=  W_i^{\rm tree}(\qqh)\, 
\delta(1+\hat{q}^2-2\hat{q}_0), $$ 
$k_+$ is the light-cone component of the residual $b$ quark momentum,
and $ F(k_+)$ is the distribution function.  Eq.~(\ref{eq:conv1}) is
valid at the leading order in $1/m_b$ and does not include
perturbative contributions.  

The main properties of the distribution function $F(k_+)$ in
Eq.~(\ref{eq:conv1}) are well-known: while it cannot be presently
computed from first principles, its moments follow from the local OPE
because they are related to the $q_0$-moments of the structure
functions $W_i$. The distribution function is universal, {\it i.e.}\
independent of the considered structure function 
and shared by inclusive radiative and
semileptonic $B$ decays. Moreover, since the leading-order moments of
$F(k_+)$ are independent of $q^2$, the distribution function does not
depend on $q^2$.

Perturbative corrections can be included in the leading twist formula
Eq.~\eqref{eq:conv1}, by using in the convolution the
short-distance perturbative structure functions that contain gluon
bremsstrahlung and virtual corrections, instead of the tree-level
$W_i^{(0)}(q_0,q^2)$. 

The phenomenological and conceptual importance of subleading
contributions to Eq.~(\ref{eq:conv1}) has been repeatedly stressed in
the last few years \cite{benson,subSF}.  In order to proceed beyond
the leading order, we first of all modify Eq.~(\ref{eq:conv1}) into
\begin{equation} \label{eq:conv1a}
W_i(q_0,q^2) = m_b^{n_i} \,\int dk_+ \ F(k_+) \ W_i^{\rm pert}
\left[ q_0 - \frac{k_+}{2} \left( 1 - \frac{q^2}{m_b M_B} \right), q^2 \right].
\end{equation}
The advantage of the latter representation is that it
automatically yields the correct hadronic endpoint for $q_0$ at
arbitrary $q^2$: since $k_+^{max} = \bar\Lambda$, the maximum value for
$q_0$ is
\begin{equation}
q_0^{max} = \frac{k_+^{max}}{2} \left( 1 - \frac{q^2}{m_b M_B} \right) +
\frac{m_b^2 + q^2}{2m_b} = \frac{M_B^2 + q^2}{2 \ M_B}.
\end{equation}
Following \cite{benson} we do not split the distribution
function into separate leading and subleading components.  Rather, we
define the finite-$m_b$ distribution function assuming the form of the
convolution in Eq.~(\ref{eq:conv1a}) to hold at any order.  

Clearly, such a distribution function is no longer universal: there
must be a distribution function for each structure function and none
of them corresponds exactly to the one describing the radiative
decays. Similarly, since their moments including power corrections
depend explicitly on $q^2$, the distribution functions depend on
$q^2$.  Finally, like the OPE parameters and the perturbative
corrections, the distribution functions depend on the infrared cutoff
$\mu$: by construction, they are designed to absorb all infrared
physics characterized by energy scales below $\mu$.  The form
\eqref{eq:conv1a} of the convolution amounts to their perturbative
definition.

  We will
therefore work with three distribution functions $F_i(k_+,q^2,\mu)$
that, after including the perturbative corrections according to
Eq.~(\ref{eq:conv1a}), must lead to $\mu$-independent distributions.
Thus, our generalized convolution reads
\begin{equation} \label{eq:conv2}
 W_i(q_0,q^2) = m_b^{n_i}(\mu)
 \int dk_+ \ F_i(k_+,q^2,\mu) \ W_i^{pert}
\left[ q_0 - \frac{k_+}{2} \left( 1 - \frac{q^2}{m_b M_B} \right), q^2,\mu \right]
\end{equation}

As mentioned already, all the available information on the
distribution functions is encoded in their moments\footnote{The photon
spectrum of $B\to X_s \gamma$ is currently measured with a good
accuracy at the $B$ factories for $E_\gamma>2.0$~GeV. These data are
important for the precise determination of $m_b$ and of the OPE
parameters.  However, the underlying distribution functions in
semileptonic and radiative decays differ at the subleading level, and
some of the subdominant effects do not seem to be under control
\cite{neubert-paz}. Therefore, we believe that the photon spectrum in
$B\!\to\! X_s \gamma$ should not be used directly to determine the
distribution function of semileptonic decays. On the other hand, as
illustrated in \cite{benson} and its phenomenological applications,
the photon spectrum can be accurately predicted, using our method, for
$E_\gamma$ not too high.}.

\begin{figure}[t]
\begin{center}
\mbox{\epsfig{file=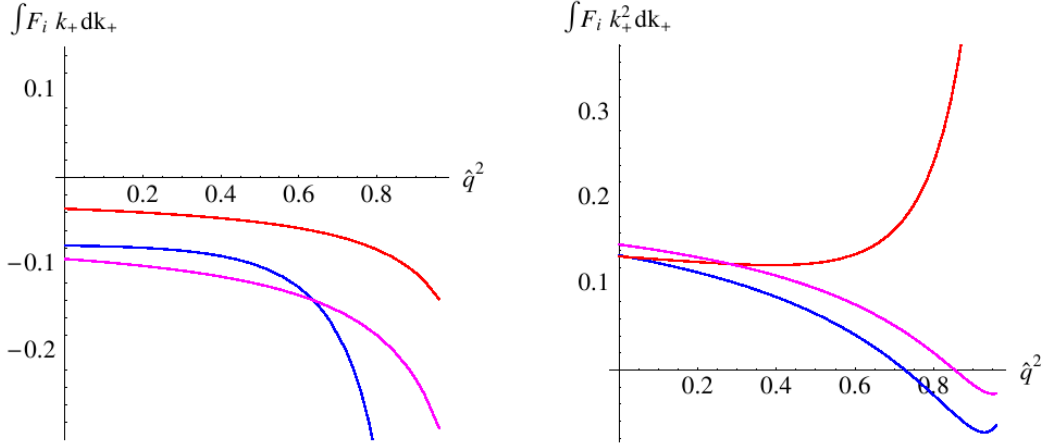,width=14cm}}
\end{center}
\caption{\sf Dependence of the first and second 
moments of the distribution functions 
$F_i(k_+,q^2,\mu)$ on $q^2$.
The blue, red, and magenta curves correspond to 
$i=1,2,3$, respectively. }\label{Fiq2}
\end{figure}

They can be extracted by matching 
with the OPE predictions for the $q_0$-moments of the structure functions, 
 known through $O(\as^2 \beta_0)$ and $1/m_b^3$. Since the perturbative
corrections to the Wilson coefficients of the power suppressed operators are 
not known, the moments of the $F_i(k_+,q^2,\mu)$ can be determined
neglecting perturbative corrections. 
 We now consider the following moments of the structure functions
 \begin{eqnarray} 
\label{SF_matching}
\int d q_0 (q_0 - a)^n \, W_i(q_0,q^2) =
m_b^{n_i} \int d k_+ \, F_i(k_+,q^2) 
 \int d q_0 (q_0 - a)^n \, W_i^{(0)}(q_0 - k_+ \frac{ \Delta}2  ,q^2),
\end{eqnarray}
where we have left all $\mu$-dependence implicit. We have also employed
\begin{equation}
a = \frac{m_b^2 +q^2}{2 \ m_b}, \quad\quad\quad
\Delta = 1 - \frac{q^2}{m_b M_B}, 
\end{equation}
and replaced $ W_i^{pert}$ with  its lowest order term $W_i^{(0)}$.
The lhs of Eq.~\eqref{SF_matching} is calculated including 
power corrections and reads
\begin{equation}
m_b^{n_i} \int d q_0 (q_0 - a)^n \left[ W_i^{(0)}(q_0,q^2)
+ W_i^{pow}(q_0,q^2)\right]= m_b^{n_i+n+1}
\left(I_i^{(n),{\rm tree}}+I_i^{(n),pow}\right),
\end{equation}
with the  $I_i^{(n)}$'s  given in the Appendix B, while 
the rhs becomes
\begin{eqnarray}
m_b^{n_i+1} \left(\frac{\Delta}{2}\right)^n \, 
I_i^{(0),{\rm tree}} \,\int d k_+ \ k_+^n \ F_i(k_+,q^2),
\end{eqnarray}
where we have used the  vanishing of all central moments of the
tree-level structure functions for $n>0$. $I_i^{(0),{\rm tree}}$ is also 
given in App.\ B.
The moments of the distribution functions are therefore given by
\begin{equation}\label{eq:SF_constraints_semilept}
\int d k_+ \ k_+^n \ F_i(k_+,q^2) = \left(\frac{2\, m_b}{\Delta}\right)^n 
\left[ \delta_{n0} + \frac{I_i^{(n),pow}}{I_i^{(0),{\rm tree}}}\right].
\end{equation}
The functions $F_i(k_+,q^2)$ are normalized to 1, up to small 
power corrections.
The $q^2$ dependence of the first two moments is shown in Figs.~\ref{Fiq2}.
The importance of subleading contributions is apparent: the first
moments for the three structure functions are quite apart from each
other and we observe a strong $q^2$-dependence at high $q^2$, where
they even diverge. The second moments of $F_{1,3}$ decrease with
increasing $\qqh$ and have a zero at $\qqh\approx 0.7-0.8$. The
variance of these two distribution functions decreases for growing
$q^2$, until it reaches negative values and the very concept of
(positive definite) distribution function becomes meaningless.  This
is not surprising: as already noted in \cite{Bigi:1993ex}, the
light-cone distribution function of the heavy quark cannot describe
the semileptonic decay at high $q^2$.  We will come back to this
subject in a dedicated section.

 It is possible to use alternative forms of the convolution of the
short and long-distance contributions, which differ at the subleading
level from Eq.~\eqref{eq:conv2}. For instance, one could use $m_b^2$
or $M_B^2$ instead of $m_b M_B$ in the argument of $W_i^{\rm
pert}$. However, this would simultaneously redefine the power
corrections to the moments of the light-cone function, and the
consequent change in the distribution functions would largely
compensate the change in the convolution at the level of observable
structure functions.  Therefore, adopting our procedure the choice of
convolution is not an ad-hoc assumption.

Once the distribution functions are required to respect the OPE
relations for their moments with power accuracy, the only element of
arbitrariness concerns the ansatz that is employed for their
functional form (see next Section).  The latter is not specific to the
treatment of the subleading-twist effects and is present already in
the leading-twist analysis. Varying the functional form will allow us
to estimate the associated theoretical uncertainty.  The convolution
\eqref{eq:conv2} can now be used to compute the structure functions.
It is useful to make the $\theta$-function contained in $F(k_+)$
explicit:
\begin{equation} \label{eq:FandG}
F_i(k_+,q^2,\mu)=G_i(k_+,q^2,\mu) \ \theta(\bar{\Lambda} - k_+),
\end{equation}
and write the detailed form of the convolution as
\begin{align}
\frac{W_i(\qz,\qq)}{ m_b^{n_i+1}} &= \frac{1}{\Delta} \Bigl[ W_i^{\rm tree}(\qqh) + 
\frac{C_F \alpha_s}{\pi} V_i^{(1)}(\qqh,\eta)+ 
\frac{C_F \alpha_s^2\beta_0}{\pi^2} V_i^{(2)}(\qqh,\eta) \Bigr] 
G_i\left(\frac{2\qzh-1-\qqh}{\Delta},\qqh,\mu \right) \nonumber\\
& - \ \frac{1}{\Delta^2} \frac{C_F \alpha_s}{\pi} \left[ 
B_{i}^{(1)}(\qqh,\eta) +\frac{\as\beta_0}{\pi} B_{i}^{(2)}(\qqh,\eta)\right]
G_i'\left(\frac{2\qzh-1-\qqh}{\Delta},\qqh \right) 
 \label{eq:W_convolution_direct_kin}
\\ &+ \frac{C_F \alpha_s}{\pi}
\int_{\frac{2\qzh-1-\qqh}{\Delta}}^{\lamh}
 d\kappa\, G_i(\kappa,\qqh,\mu) 
\left[
R_i^{(1)}\left(\qzh-\frac{\Delta}{2}\kappa,\qqh \right) +
\frac{\alpha_s\beta_0}{\pi}
R_i^{(2)}\left(\qzh-\frac{\Delta}{2}\kappa,\qqh \right) \right] \nonumber
\end{align}
Here we have employed $\lamh=\bar\Lambda/m_b$.
\begin{figure}[t]
\begin{center}\hspace{-10mm}
\mbox{\epsfig{file=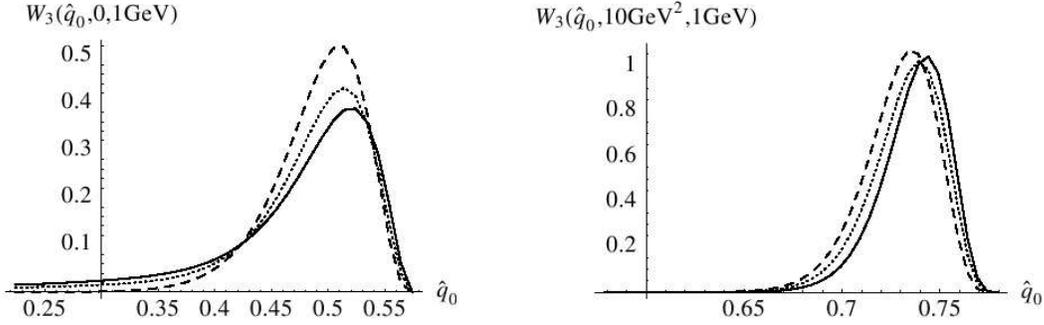,width=14cm}}
\end{center}
\caption{\sf The convoluted $W_3$ 
at $q^2=0$ (left) and $q^2=10\,{\rm GeV}^2$ (right) for  $\mu=1$~GeV.
The dashed, dotted, and solid curves  correspond to 
lowest order, next-to-leading order, and $O(\as^2 \beta_0)$ 
in the perturbative corrections, respectively. }\label{fig:convff}
\end{figure}
In Fig.~\ref{fig:convff} we show the $q_0$-dependence of the
convoluted $W_3$ at $q^2=0$ and $q^2=10\,{\rm GeV}^2$ for $\mu=1$~GeV
for the exponential ansatz discussed in the next section. The plots
compare the structure function calculated at the leading order (dashed
curves) and at the next-to-leading order (dotted curves) in $\as$, and
including also the $O(\as^2 \beta_0)$ corrections (solid curves).  We
observe that the width of the convoluted structure function shrinks
significantly between $q^2=0$ and $q^2=10\,{\rm GeV}^2$.  The main
effect of perturbative corrections at small $q^2$ is a broadening of
the shape, due to real gluon emission.  At higher $q^2$, instead, real
emission is progressively inhibited.  In the example considered,
$q^2=10\,{\rm GeV}^2$ lies just below $(1-2\eta)\, m_b^2$ (see
Fig.~\ref{q0-q2}). Therefore the only appreciable perturbative effect
is a shift of the peak somewhat towards higher $q_0$ values, and is
driven by the $\delta'$ term in Eq.~\eqref{cutSp}.  The $W_{1,2}$
structure functions behave in a very similar way.

\section{Functional forms}
We have seen that only the first few moments of the distribution
functions are known. They are given in terms of matrix elements of
local operators that are measured in $b\to c$ semileptonic and
radiative decays.  The modelling of QCD dynamics in the threshold
region therefore requires an {\it ansatz} for the functional form,
which must comply with the constraints coming from the first few
moments.  Several functional forms have been proposed in the
literature. They have to be exponentially suppressed at large negative
$k_+$ and to vanish at the endpoint $k_+=\bar\Lambda$. We will further
assume the positivity of $F_i(k_+)$. As suggested by its probabilistic
interpretation, the primordial $F_i(k_+)$ should be positive definite,
but radiative corrections can potentially change its sign, in analogy
with the parton distribution functions of deep inelastic scattering.
In our approach with the Wilson cutoff $\mu$ effects of this kind are
excluded in the negative tail and we choose to neglect them
altogether, assuming positive distribution functions.

The basic two-parameter functional forms proposed in the 
literature \cite{Dikeman:1995ad,BLNP,benson} are  
\bea\label{basic1}
F(k_+)&=& N \,(\bar{\Lambda}-k_+)^a \, e^{b \,k_+}\
 \theta(\bar\Lambda-k_+)\quad\quad \quad\quad
{\rm (exponential)}\\
F(k_+)&=& N\,(\bar{\Lambda}-k_+)^a\, e^{-b \,(\bar\Lambda-k_+)^2} \, \theta(\bar\Lambda-k_+)
 \quad\quad{\rm (gaussian)}\\
F(k_+)&=& N\,\frac{(\bar{\Lambda}-k_+)^a}{\cosh \left[b (\bar{\Lambda}-k_+)\right]}\, \theta(\bar\Lambda-k_+)
 \quad\quad\quad\ {\rm (hyperbolic)}\\
F(k_+)&=& N e^{-a \left(\bar\Lambda -k_+ + \frac{b}{\bar\Lambda -k_+}\right)^2}\, \theta(\bar\Lambda-k_+)\quad\quad\quad\quad {\rm (roman)}\label{basic2}
\eea
\begin{figure}[t]
\begin{center}
\mbox{\epsfig{file=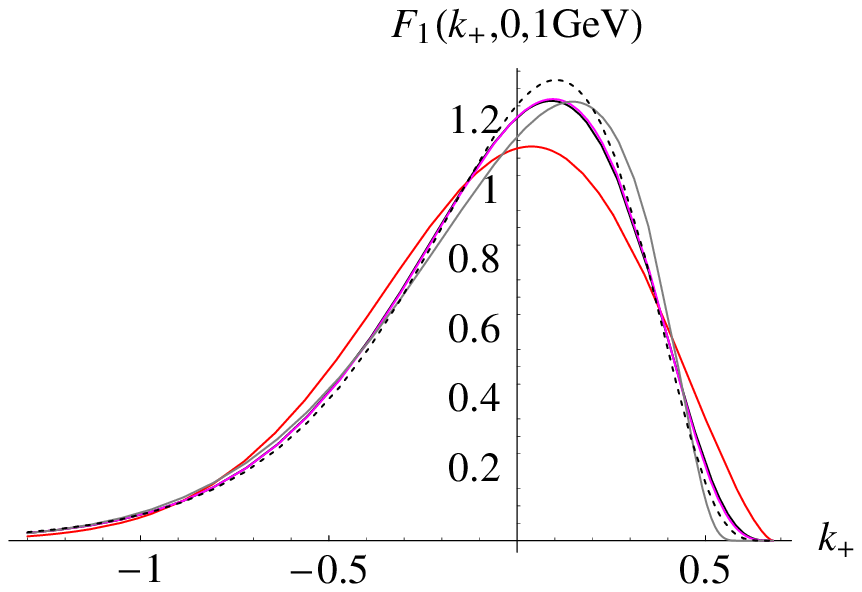,width=8cm}}
\mbox{\epsfig{file=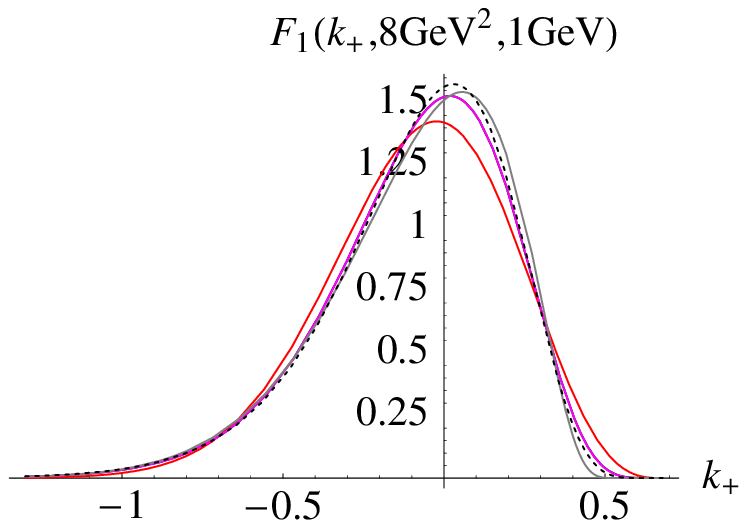,width=7.5cm}}
\end{center}
\caption{\sf A comparison of the basic functional forms in 
Eqs.~(\ref{basic1}-\ref{redexp}) for the distribution function $F_1(k_+,q^2,\mu)$
after using the first two moments at $q^2=0$ and $q^2=8$GeV$^2$. }\label{fig:basic}
\end{figure}
The parameters $a,b>0$  are fixed by the 
first two normalized moments of $F(k_+)$ and can be 
easily found numerically. 
A comparison of the  resulting distribution functions $F_1(k_+,q^2,\mu)$
at $q^2=0$ and $q^2=8$~GeV$^2$ is shown in Fig.~\ref{fig:basic}.
The numerical inputs for the moments are taken from \cite{BF}.
The results for the other two light-cone functions are very similar.
The hyperbolic and exponential forms are 
almost indistinguishable in the plot. Indeed, once the first two moments are 
fixed, the shape of the various curves 
is determined to a large extent by the properties of the tails. In this 
respect, there is not a large variety in  \eqref{basic1}-\eqref{basic2}. 
It is not difficult to find alternatives:
for instance, one could modify the exponential form as in 
\be\label{redexp}
F(k_+)= N \,(\bar{\Lambda}-k_+)^a \, e^{b \,(\bar{\Lambda}-k_+)^{\frac23}}\, \theta(\bar\Lambda-k_+), 
\ee
which is also displayed in Fig.~\ref{fig:basic} (dashed curve) and is 
characterized by a higher negative tail.
\begin{figure}[t]
\begin{center}
\mbox{\epsfig{file=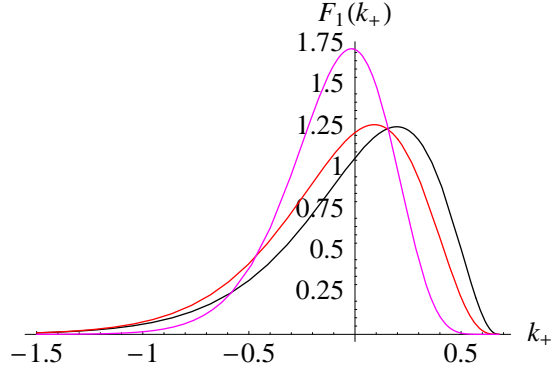,width=8cm}}
\end{center}
\caption{\sf Exponential form for $F_1(k_+,q^2,\mu)$
retaining only the leading contributions to its moments (black curve)
and including all known subleading effects at $q^2=0$ (red)
and at $q^2=10\GeV^2$ (magenta).}
\label{fig:sublead}
\end{figure}
It is  interesting to see the effect of subleading corrections to the 
moments of the distribution functions: this is illustrated in 
Fig.~\ref{fig:sublead}, where  $F_1(k_+,q^2,\mu)$ is computed using the
exponential  ansatz retaining only the leading contributions to its 
moments, and including all the known power corrections.

The need for more flexibility in the choice of the functional form can
also be understood by comparing the third normalized moments of the
curves displayed in Fig.~\ref{fig:basic}. For a given value of $q^2$,
they are remarkably close to each other. Let's consider $q^2=0$: while
the gaussian form yields $-0.054$ for the third normalized moment,
exponential, hyperbolic, and roman forms cluster around $-0.071$, and
that in \eqref{redexp} yields $-0.079$. On the other hand, the OPE
predicts the third moment to be $-\rd/3$, up to unknown $O(1/m_b)$ and
$O(\as)$ corrections.  Using the results of the fit in \cite{BF}, this
amounts to $-0.058\pm 0.007$, in rough agreement with the values
mentioned above.  The situation does not change significantly with
$q^2$, but the spread between different forms decreases further.

Clearly, a mismatch between the third moment of the two-parameter
forms we have considered and the OPE prediction might signal either
large and possibly $q^2$-dependent subleading contributions to the
OPE, or that the specific functional forms are disfavored by present
data on inclusive semileptonic moments. A line is difficult to draw,
but none of the basic forms in (\ref{basic1}--\ref{redexp}) appears
particularly disfavored.  Rather than using the third moment to
constrain the distribution functions, we have used it to gauge 
the diversity  in the functional forms we employ.

\begin{figure}[t]
\begin{center}
\mbox{\epsfig{file=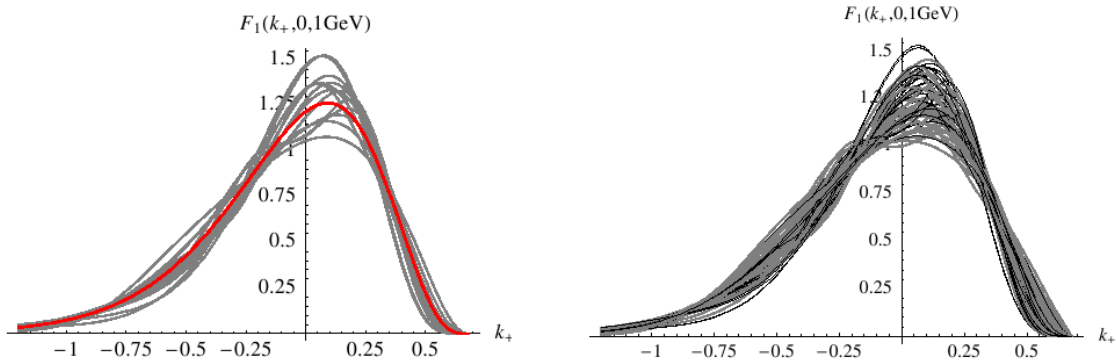,width=15cm}}
\end{center}
\caption{\sf Left: a sample of possible ways to distort  
the exponential form (red curve) in 
 Eq.~(\ref{basic1}) compatible with the first two moments
for the  distribution function $F_1(k_+,0,1{\rm GeV})$. Right: the same 
but considering other basic functional forms.}\label{fig:distorsion}
\end{figure}
A straightforward generalization of the above basic forms consists in 
multiplying  them by a distortion factor $D (k_+)$ that can be
a polynomial, positive in the  $(-\infty,\bar\Lambda)$ domain,
or a more general positive function. In principle the distortion
can depend on $q^2$. 
Of course not all the distortions $D(k_+)$ yield acceptable
solutions. In particular, we discard shapes with two or more maxima,
but keep those with an inflection point. We are particularly
interested in distortion factors that modify the third moment. This
can happen when the distortion enhances or suppresses the lower
(negative $k_+$) tail, or, to a lesser extent, with an asymmetric and
bounded function, like a constant with a sinusoidal perturbation. In
general it is difficult to find distortions that modify significantly
the third moment.  A representative but not exhaustive sample of
distribution functions $F_1(k_+,q^2=0)$, based on the exponential form
and complying with the first and second moment constraints, is shown
in Fig.~\ref{fig:distorsion}.  All the functions in the sample satisfy
the first two moment constraints and yield third moments that differ
by less than 30\% from the basic exponential form. Since there is no
reason of principle to discard any of them, they will be used,
together with similar ones, in our study of the theoretical
uncertainty. Changing also the basic functional form one gets the plot
on the rhs of Fig.~\ref{fig:distorsion}, with almost a hundred
different forms.

\section{The high-$q^2$ tail}
We have seen that the formalism developed in Sec.~3 cannot be applied
at high $q^2$. The relevance of Fermi motion subsides at high $q^2$
and Fig.~\ref{Fiq2} shows that the variance of the distribution
functions $F_{1,3}$ becomes negative at $\qqh\sim 0.7$-0.8, as a
consequence of $O(\Lambda^3)$ effects. While the leading contributions
to the $n$-th central moment of the structure function $W_i$,
$I_i^{(n)}/I_i^{(0){\rm tree}}$, is suppressed by $(1-\qqh)^n$ at large
$\qqh$, subleading contributions in $1/m_b$ have a weaker suppression
that enhances their weight at high $q^2$.

We also recall that our discussion of perturbative corrections points
to larger uncertainties in the high $q^2$ region. We have seen in
Sec.~2 that the expressions for the  perturbative corrections become
non-analytic in $\eta=\mu/m_b$ for $\qqh> (1-2\eta)^2$.  This is a
manifestation of the same problem mentioned above: at high $q^2$ the
contribution of higher dimensional operators is no longer suppressed.

\begin{figure}[t]
\begin{center}
\mbox{\epsfig{file=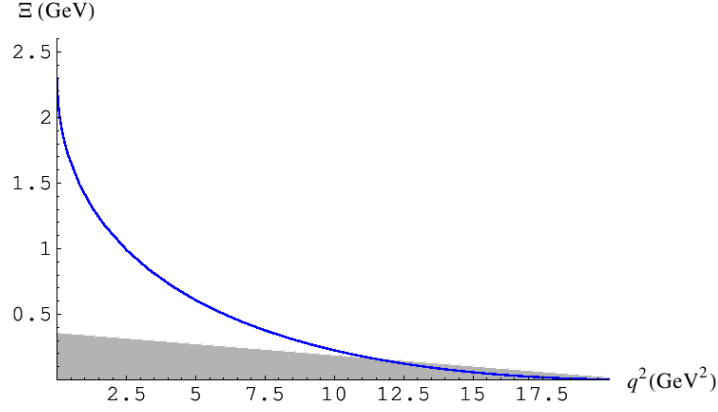,width=10cm}}
\end{center}
\caption{\sf The width $\Xi$ of the $q_0$ range expressed in GeV 
as a function of $q^2$ for $m_b=4.6$~GeV. The 
gray band corresponds to  $\Xi\le\sqrt{\mu_\pi^2 /3}(1-\qqh)$ 
(see text). }\label{fig:Xi}
\end{figure}

A simple way to visualize the problem is to plot the width of the
physical range in $q_0$ as a function of $q^2$, as in
Fig.~\ref{fig:Xi}.  The OPE structure functions are smeared over a
region in $q_0$ of width $\Xi=(m_b -\sqrt{q^2})^2/2m_b$.
$\Xi$ shrinks rapidly with $q^2$, reaching a size comparable to
$\Lambda_{\rm QCD}$ for $q^2\sim 10\GeV^2$.  The range $\Xi$ can also be
compared with the typical width of the structure function, which at
the leading order in $1/m_b$ is
$\sqrt{\frac{\mu_\pi^2}3}(1-\qqh)\approx 0.36\,(1-\qqh)$~GeV.  As
shown in Fig.~\ref{fig:Xi}, they are equal for 
$q^2\sim 11\GeV^2$
and for  higher $q^2$ the width of the physical $q_0$-range is
smaller than the width of the structure functions, 
a situation that makes all local OPE results unreliable.  


It is indeed well-known that for very large $q^2$ the local OPE fails
to provide a reliable description of the semileptonic decay, as the
dynamics in that region is not characterized by a short-distance scale
\cite{Bigi:1993ex,WA}.  The Wilson coefficients in the power
expansion, for instance, grow like $1/(1-\qqh)^n$.  The origin of this
enhancement  at high $q^2$ can be understood
by calculating the $q^2$-differential rate as the integral over $q_0$ of the
double differential rate,
\be \frac{d\Gamma}{d\qzh \, d\qqh}\sim
\sqrt{\qzh^2-\qqh} \left\{ 3 \qqh \,W_1+ (\qzh^2-\qqh)\, W_2\right\}.
\label{double}
\ee 
It is related to the presence of the square root, a non-analytic term
that follows from the kinematic constraints on the electron energy ---
the first $\theta$-function in Eq.~\eqref{eq:aquila_normalization}.
Expanding the square root in $\qzh$ around the endpoint
$\qzh=(1+\qqh)/2$ one gets 
\be \sqrt{\qzh^2-\qqh} = \frac{1-\qqh}2
-\sum_{n=1}^\infty \frac{(-2)^{n} \, b_n(\qqh)}{(1-\qqh)^{2n-1}}
\left(\qzh-\frac{1+\qqh}2\right)^n\label{series}.  \ee
Clearly, the convergence radius of the series gets smaller at higher 
$\qqh$ and,  in particular, higher orders become more and more important.
Replacing the square root in \eqref{double} with its expansion 
\eqref{series} and integrating over $\qzh$ one sees that non-perturbative 
contributions are organized in the series
\be
\frac{d\Gamma}{dq^2}\sim -\sum_{n=1}^\infty \frac{(-1)^n \, b_n (\qqh)}{(1-\qqh)^{n-2}}
\left(\frac{\bar\Lambda}{m_b}\right)^n, 
\label{q2sing}
\ee 
and one can verify that unknown $O(1/m_b^4)$ corrections 
become comparable to $O(1/m_b^3)$ corrections at $\qqh\sim 0.7$.

Since the contributions of  higher dimensional operators 
become more and more singular at high $q^2$, 
we see from \eqref{q2sing} that their contribution to 
 the $b\!\to\!q\,\ell\nu$ total width is singular for $n\ge 3$. 
In particular, the Darwin operator contribution to the 
$q^2$ differential rate is ($\qqh\neq 1$)
\be
\frac{d\Gamma}{d\qqh}\sim  \frac{\rd}{6m_b^3}\left[
20\, {\hat q}^6 +66\, {\hat q}^4+48\, \qqh+74 -\frac{96}{1-\qqh}\right]+...
\label{q2ope}
\ee 
which upon integration over $q^2$ leads to a logarithmic
singularity that can be regulated by the quark mass $m_q$
\cite{1mb3,bds}. This unphysical singularity is removed by a one-loop
penguin diagram that mixes at $O(\as)$ the Weak Annihilation
four-quark operator $O_{\scriptscriptstyle\rm WA}^u=-4 \bar
b_L^\alpha\vec\gamma u_L^\alpha\,\bar{u}_L^\beta\vec\gamma b_L^\beta$
into the Darwin operator.  The cancellation is discussed in detail in
\cite{bumoments}. Defining \be B_{\scriptscriptstyle\rm WA}\equiv
\langle B|O_{\scriptscriptstyle\rm WA}^u|B\rangle, \quad\quad
C_{\scriptscriptstyle\rm WA}=32\pi^2 / m_b^3, \ee
one finds that the sum of the contributions of the WA and Darwin
operators to the semileptonic total width is
\be
 \delta \Gamma\sim \left[C_{\scriptscriptstyle\rm
WA}\, B_{\scriptscriptstyle\rm WA}(\muwa) -
\left(8\ln\frac{m_b^2}{\muwa^2}-\frac{77}{6}\right)
\frac{\rho_D^3}{m_b^3} +{\cal O}(\as)\right],
\label{wa}
\ee
which is free of singularities.  The scale $\muwa$ is the
$\overline{\rm MS}$ renormalization scale of the WA operator.  The
constant accompanying the logarithm in Eq.~(\ref{wa}) depends on the
renormalization scheme; we have employed here the same scheme as in
Ref.~\cite{bumoments}.  At the level of the $q^2$ differential spectrum
of Eq.~\eqref{q2ope}, the singularity $1/(1-\qqh)$ is 
replaced by $1/(1-\qqh)_+$ and is accompanied by a term 
 $\delta(1-\qqh)$ whose coefficient contains
 $B_{\scriptscriptstyle\rm WA}$ and can be read off
directly from  Eq.~(\ref{wa}).

It is well-known that in the factorization approximation the matrix
element $B_{\scriptscriptstyle\rm WA}$ vanishes, and that WA is
phenomenologically important only to the extent factorization is
actually violated. Since Eq.~\eqref{wa} is independent of the scale
$\muwa$, up to $O(\as)$ corrections that we neglect, we have in
the $\overline{\rm MS}$ scheme $B_{\scriptscriptstyle\rm
WA}(\mu')=B_{\scriptscriptstyle\rm WA}(\mu)- \rd/2\pi^2 \, \ln \mu'/\mu$.
Clearly factorization may hold only for a certain
value  $\muwa=\mu_f$ and, therefore, if factorization holds at 
$\mu_f$, namely $B_{\scriptscriptstyle\rm WA}(\mu_f)=0$, a
change of the scale $\mu_f$ provides a rough measure of the (minimal)
violation of factorization induced perturbatively.

\begin{figure}[t]
\begin{center}
\mbox{\epsfig{file=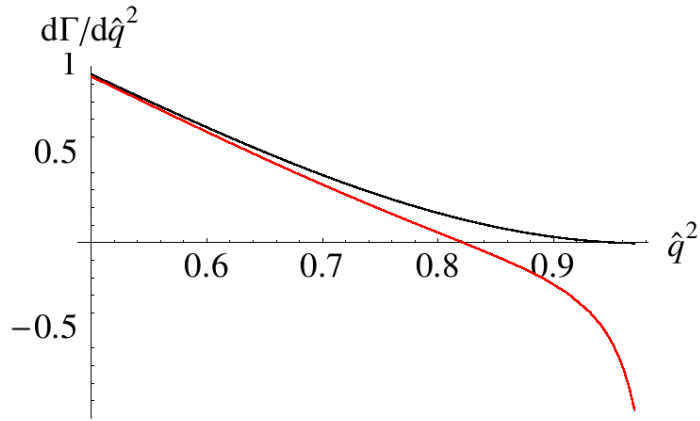,width=10cm}}
\end{center}
\caption{\sf The $q^2$ spectrum in the endpoint region with perturbative 
corrections switched off. The red curve follows 
from the OPE prediction, the black one modifies it according to 
Eq.~(\eqref{q2mod}) to guarantee
positivity ($b=0.4$).  }\label{fig:highq2}
\end{figure}

We should stress that the inclusion of WA removes the unphysical
singularity in the $1/m_b^3$ OPE but not its intrinsic limitations at
high $q^2$.  The local OPE non-perturbative contributions for the rate
with arbitrary cuts on $M_X$, $E_\ell$, and $q^2$ is reported in
Appendix C.  It includes WA contributions as discussed above.  The
main spectra follow from these expressions upon differentiation.  It
can be easily seen that both the rate and some of the differential
spectra become negative for a sufficiently high cut on $q^2$, a clear
sign of the importance of higher dimensional operators.  This is
apparent in the case of Eq.~\eqref{q2ope} where the $1/(1-\qqh)$
singularity drives the $q^2$ spectrum to negative values for $q^2>
17.5\GeV^2$, see Fig.~\ref{fig:highq2}.\footnote{This is by far the
dominant but not the only negative contribution. Even in the absence
of the Darwin term the differential rate becomes negative at very high
$q^2$.}

Because of the limitations of the OPE in the high $q^2$ region, it is
preferable to model the $q^2$ tail in a way that is both compatible
with the positivity of the spectra and incorporates some features of
the OPE. We want to modify the OPE expressions before the differential
rate becomes negative. To this end we change approach for $q^2$
greater than a certain $q_*^2$ and adopt one of the two following
methods.

The first method is based on the local OPE result without
the convolution with the light-cone distribution functions.  One starts
from the OPE distribution \eqref{q2ope} and introduces a damping
factor in the $1/(1-\qqh)$ singularity 
\be \frac{d\Gamma}{d\qqh}\sim\left\{
\frac{\rd}{6m_b^3}\left[ 20\, {\hat q}^6 +66\, {\hat q}^4+48\, \qqh+74
-\frac{96\,(1-e^{-\frac{(1-\qqh)^2}{b^2}} )} {1-\qqh}\right]+ X\,
\delta(1-\qqh)+...\right\}
\label{q2mod}
\ee
that maintains the $q^2$ differential rate positive for appropriate
values of $b$ (see Fig.~\ref{fig:highq2}).  We have also explicitly
written the Dirac delta at the endpoint.  The damping factor can also
be applied to the $E_\ell$ and $M_X$ differential distributions.  In a
realistic setting one could expect a smooth (positive) bump close to
maximal $q^2$. However, since in the following we will always
integrate over the $q^2$ tail, the rough modelling of \eqref{q2mod} is
sufficient.  It should be clear that positivity of the rate implies
$X\ge 0$. The dimensionless parameter $X$ is related to the WA matrix
element and to the WA scale $\muwa$ by integrating over
$q^2>q_*^2$.\footnote{  The  relation is $
X=C_{\scriptscriptstyle\rm WA} B_{\scriptscriptstyle\rm WA}(\muwa)
-8\frac{\rd}{m_b^3}\left[ 1-\gamma_E+{\rm Ei}
\left(-\frac{(1-\qqh_*)^2}{b^2}\right)+2 \ln \frac{b\,
m_b}{\muwa}\right] $.}  In the case of our default choice $b=0.4$, the
minimum value $X=0$ corresponds to  $B_{\scriptscriptstyle\rm
WA}(1\GeV)\sim 0.008\GeV^3$ or equivalently to a factorization scale
for WA $\mu_f\sim 2.2\GeV$.  The amount of WA implied by this
ansatz for the high-$q^2$ tail grows with $X$: $X=0.03$ corresponds, 
in the default
setting, to $B_{\scriptscriptstyle\rm WA}(1\GeV)\sim 0.017\GeV^3$ 
($\mu_f\sim 5.7\GeV$).  Conversely, the experimental determination of
$B_{\scriptscriptstyle\rm WA}(1\GeV)$ would allow us to fix the value of
$X$. 

Since we will mainly be interested in integrated quantities (for
instance total rates with cuts), this approach is based directly on
the formulas in App.~C and does not provide a triple differential
distribution in the high $q^2$ region, but includes the endpoint
effect parameterized by $X$.  Perturbative corrections are implemented
as for $q^2<q^2_*$ but again there is no convolution.  We adopt this
first method as our default choice.

In the second method that we employ to model the high $q^2$ region we
freeze the distribution functions at $q^2=q^2_*$ and use it in the
convolution formula for all $q^2>q^2_*$.  Because of the decreasing
phase space available at higher $q^2$, this approach effectively
amounts to gradually phasing all power-corrections out.  Although WA
does not explicitly enter in this case, the ansatz corresponds to a
certain amount of WA depending on the value of $q_*^2$ and on the
precise values of the other heavy quark parameters.  For instance,
for $q^2_*=11\GeV^2$ and the default values of the non-perturbative
parameters, it corresponds to $B_{\scriptscriptstyle\rm WA}(1\GeV)\sim
0.001\GeV^3$ ($\mu_f\sim 1.1\GeV$).  An additional amount of WA can
be easily accommodated in this framework as an extra contribution at
the $q^2$-endpoint.  The triple differential rate here is available
for each point in the phase space.

While there is no doubt that the local OPE expansion fails at high
$q^2$, the precise determination of its range of applicability is an
open question.  We employ the formulation based on the convolution of
Eq.~\eqref{eq:W_convolution_direct_kin} for $q^2<q^2_*$ with \be
8.5\GeV^2 \le q^2_*\le 13.5\GeV^2,
\label{qstar}
\ee 
keeping the maximum value of $q_*^2$ lower than the value for
which the variance of one of the distribution functions gets negative,
$q^2\sim 15\GeV^2$, because of numerical instabilities in dealing with
very narrow light-cone functions.  To model the high-$q^2$ tail at
$q^2>q^2_*$ we employ one of the two methods described above.  The
difference of the two approaches, as well as a variation of the
parameters in Eqs.~(\ref{wa},\ref{q2mod},\ref{qstar}) provide us with
an estimate of the theory uncertainty in the high $q^2$ region.

\section{Results and theoretical errors}
\begin{figure}[t]
\begin{center}
\mbox{\epsfig{file=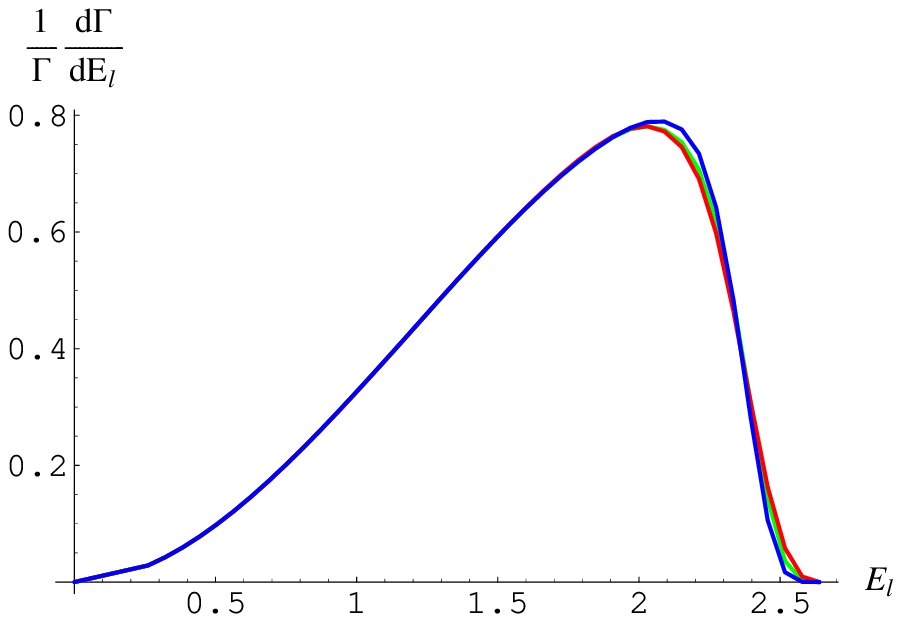,width=8cm}}
\mbox{\epsfig{file=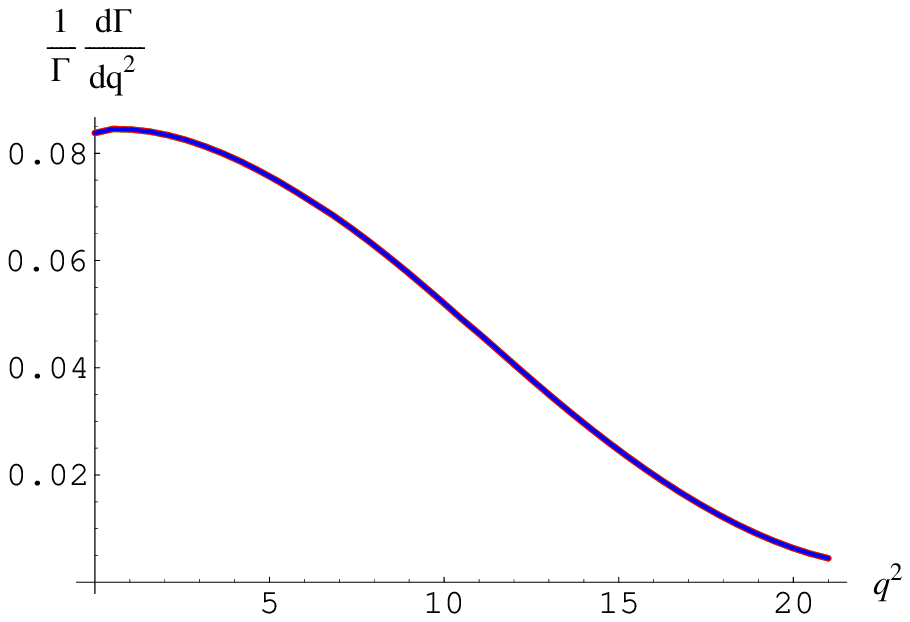,width=8cm}}
\end{center}
\caption{\sf 
Electron energy (left) and $q^2$ (right) spectra using a few different
functional forms. }\label{fig:spectra1}
\end{figure}
Let us now illustrate our method with a few applications. We start by
showing the main physical distributions following our default approach,
based on Eq.~\eqref{eq:W_convolution_direct_kin} for $q^2<11\GeV^2$,
and on the local OPE modified as in \eqref{q2mod} for higher
$q^2$.  We employ the central values of the fit in \cite{BF} for the
non-perturbative parameters at $\mu=1\GeV$, namely
\bea
&& m_b=4.613\GeV \quad\quad 
\mupi=0.408 \GeV^2 \quad\quad \mug=0.261\GeV^2\nonumber\\
&& \ \ \ \ \ \ \ \ \ \rd=0.191\GeV^3\quad\quad
\rls=-0.195\GeV^3\nonumber
\eea
%
and set $b=0.4$ and $X=0$
in Eq.~\eqref{q2mod}.  All numerical results in this section are obtained
with a C++ implementation of the procedure described in the previous
sections.

\begin{figure}[t]
\begin{center}
\mbox{\epsfig{file=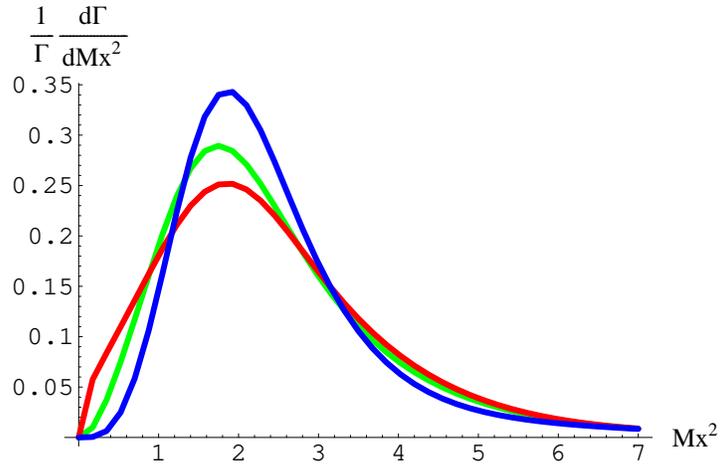,width=10cm}}
\end{center}
\caption{\sf 
Invariant hadronic mass spectrum using three different 
functional forms. }\label{fig:spectra2}
\end{figure}

Fig.~\ref{fig:spectra1} shows the electron energy and $q^2$ spectra
using a few different functional forms. Since we impose the local OPE
constraints on the distribution functions at fixed $q^2$, the $q^2$
spectrum is by construction independent of the adopted functional
form, while the lepton energy spectrum shows a limited dependence.
The hadronic invariant mass spectrum is displayed in
Fig.~\ref{fig:spectra2}. In this case the difference between
functional forms is more pronounced.

One of the checks we have performed concerns
the total width that in our scheme  is given by
\begin{eqnarray}
\Gamma_{tot}& =& 
\Gamma_0\left(1 +2\, \frac{\alpha}{\pi} \,\ln \frac{M_Z}{m_b}\right)
\left[ 1
+ C_F \frac{\as}{\pi}\left( \frac{25}{8} - \frac{\pi^2}{2} +
\frac{20}3 \frac{\mu}{m_b}+3\frac{\mu^2}{m_b^2}-
 \frac{77}9 \frac{\mu^3}{m_b^3}
\right)  \right.\nonumber\\&&
+ C_F \ \beta_0 \left(\frac{\as}{\pi}\right)^2
\left(\frac{1009}{384}-\frac{77\pi^2}{288}-2\zeta_3 
+\frac{10 (3 \lambda+8) \mu }{9 m_b}
+\frac{(6 \lambda+13) \mu^2}{4 m_b^2}
-  \frac{77(\lambda+2) \mu^3}{18 m_b^3}
\right) \nonumber\\
&&\left.- \frac{ \mu_{\pi }^2(\mu)}{2 m_b^2}
 -\frac{ 3\mu _G^2(\mu)}{2 m_b^2} +  
\left( 77 + 48 \ln\frac{\muwa^2}{m_b^2}\right)
\frac{\rho _D^3(\mu)}{6 m_b^3} 
+\frac{3\rho _{LS}^3(\mu)}{2 m_b^3} + \ C_{\scriptscriptstyle\rm
WA}\, B_{\scriptscriptstyle\rm WA}(\mu_{\scriptscriptstyle\rm WA})  \right],
\label{total}
\end{eqnarray}
where $\Gamma_0= |V_{ub}|^2 G_F^2 m_b^5(\mu)/192\pi^3$, $\lambda= \ln
m_b/2\mu$, and we have left the $\mu$-dependence of $m_b$
implicit. The first parenthesis contains the dominant short distance 
electroweak correction \cite{sirlin} which amounts to about 1.4\%.
The perturbative QCD corrections in \eqref{total} have been
known for a while (see \cite{Gambino:2006wk} for a complete list of
references).  The non-BLM $O(\as^2)$ contributions to the total rate
are also known \cite{ritbergen} and amount to about +0.5\% in our
scheme, however we do not report them in \eqref{total} as they are
absent in our triple differential rate.  The total semileptonic decay
rate can also be obtained by integrating the triple differential
distribution computed in our default setting with the modified
high-$q^2$ tail. In this case the result is about 1.4\% higher than
Eq.~\eqref{total}, using the same input values.  This small difference is
related to the perturbative corrections at the $q^2$-endpoint and  might
in principle be accounted for by a more accurate treatment of the WA
contribution, but it will be added to theoretical error budget. Using
the second option discussed at the end of the previous section, namely
freezing the distribution functions without taking into account any
extra effect at the $q^2$ endpoint, the total rate differs from
\eqref{total} with default values by $-0.8\%$.

The next application we consider is the extraction of 
$|V_{ub}|$ from some of the latest experimental results. We consider the 
following experimental results:
\begin{itemize}
\item [\bf A] Belle analysis with $M_X\le1.7\GeV$ and $E_\ell>1.0 \GeV$ \cite{belle1}; 
\item  [\bf B] Belle and Babar analyses with  $M_X\le1.7\GeV$, $q^2>8\GeV^2$, 
and $E_\ell>1.0 \GeV$ \cite{belle1,babar2};
\item [\bf C] Babar with $E_\ell>2.0\GeV$ \cite{babar1} 
\end{itemize}
This list is far from being complete and it is meant for the purpose
of illustration only. The analysis {\bf B} was proposed in \cite{q2mx}
and suffers most from the high-$q^2$ uncertainty discussed in the
previous section (see also Ref.~\cite{uses}), but in all three cases 
the high $q^2$ region is
probed.  In comparing with experiment we avoid averaging
the various experimental results. In the case {\bf B}, however, we
take the arithmetic mean of the two experimental central values as
reference value. The partial $B\to X_u \ell\nu$ rate
given by the experiments is compared with our
theoretical predictions, yielding the values for $|V_{ub}|$ listed in Table 1: 
\be
|V_{ub}|=\sqrt{ \frac{\Gamma_{cuts}^{exp}}{
 \frac1{|V_{ub}|^2}\int_{cuts}  \frac{d^3 \Gamma_{th}}{dq_0\, dq^2\, dE_\ell}}}
\ee

\begin{table}[t]
\center{
\begin{tabular}{|c|c|c|c|c|c|c|c|c|c|c|}
\hline
cuts & {\small $|V_{ub}|\times 10^3$} & $f$  & exp & par &  pert & {\small
tail model}
& $q_*^2$ & $X$ & ff &tot th\\ \hline
{\bf A}  \cite{belle1} & 3.87 & 0.71 & 6.7  & 3.5 & 1.7 & 1.6 &
2.0  &$_{-2.7}^{+0.0}$ &$^{+2.4}_{-1.1}$ & $\pm 4.7^{+2.4}_{-3.8}$   \\
\hline
{\bf B}  \cite{belle1,babar2} & 4.44 & 0.38 & 7.3 & 3.5 & 2.6 & 3.0 & 4.0
& $_{-5.0}^{+0.0}$ & $^{+1.4}_{-0.5}$ & $\pm 6.6_{-5.5}^{+1.4}$  \\ \hline
{\bf C}  \cite{babar1} & 4.05 & 0.30 & 5.7 & 4.2 &3.3  & 1.8 & 0.9
&$_{-6.2}^{+0.0}$  &  $^{+1.2}_{-0.7}$ & $\pm 5.7^{+1.2}_{-6.9}$  \\
\hline
\end{tabular}}
\caption{\sf Values of $|V_{ub}|$ obtained using different experimental results
 and their experimental and theoretical 
uncertainties (in percentage) due to various sources (see text). 
$f$ is the estimated fraction of events. }
\end{table}
The central values given in Table~1 refer to our default setting, with an 
exponential ansatz for the distribution functions. 
The total theoretical errors in the last column of Table 1 are
obtained by combining all theory errors in quadrature, with the exception of
the asymmetric errors due to the functional forms and to $X$ (see below), which
are added linearly and kept asymmetric.

We now consider the uncertainty of our theoretical predictions for
Table~1.  There are both parametric uncertainties for instance due to
the OPE parameters ($m_b$, $\mupi$, etc.) and intrinsic uncertainties,
related to various limitations of our approach. We identify the
following sources of uncertainty:
\begin{enumerate}
\item  the value of  $\alpha_s$, 
$m_b$ and of the other non-perturbative parameters;
\item  higher order  perturbative and non-perturbative corrections;
\item  the functional form of  the distribution functions;
\item  WA and the high-$q^2$ tail.
\end{enumerate}

For what concerns the parametric errors (par), we employ $\as(m_b)=
0.22\pm 0.02$ and take the values of the non-perturbative parameters
from the fit \cite{BF} including all correlations.  Not surprisingly,
the by far dominant parametric error is related to the uncertainty in
the $b$ quark mass.  The value of $|V_{ub}|$ extracted is quite
sensitive to the precise value of the $b$ quark mass. For instance, if
one employs $m_b=4.677\GeV$, as suggested by a fit to charmed
semileptonic only, instead of $m_b=4.613\GeV$, the central values in
Table 1 change to $3.65,\,4.21,\,3.79 \times 10^{-3}$ in the {\bf
 A,B,C} cases, respectively..

 To estimate higher order perturbative corrections we $i)$ change the
hard cutoff in the range $0.7<\mu<1.3\GeV$ and $ii)$ rescale the $\sim 1\%$
discrepancy in the total rate due to perturbative effects in the
highest $q^2$ region to the fraction of events. We also consider the
overall size of the $O(\as^2\beta_0)$ corrections in our analysis
($-4.2,-6.0,-6.2$\% in the {\bf A,B,C} cases, respectively).  We take
as overall perturbative error in each case the maximum between 40\% of
the $O(\as^2\beta_0)$ corrections and the uncertainty obtained by
combining the above $i)$ and $ii)$ errors.  To estimate higher order
non-perturbative corrections and missing perturbative corrections to
the Wilson coefficients of power-suppressed operators, we vary the
non-perturbative parameters within 30\% of their central values in an
uncorrelated way. This leads to errors that are negligible in
comparison with those in Table~1.

The uncertainty due to the functional form (ff) of the distribution
functions is estimated by comparing a total of about 30 different
forms. We display in Table 1 the maximal positive and negative
variations wrt to the default exponential ansatz.  In all cases it
amounts to a relatively small error.

 Concerning the high-$q^2$ tail we consider three related sources of
uncertainty: the modelling of the $q^2$ tail, the arbitrariness in the
choice of the scale $q_*^2$ where the modelling sets it, and the WA matrix
element.  We estimate the error on modelling the $q^2$ tail by
comparing our default approach with the second method outlined at the
end of the previous section, by varying the parameter $b$ in
\eqref{q2mod} and change $q_*^2$ between 8.5 and $13.5\GeV^2$.

In connection to WA we observe that all we presently know about
$B_{\scriptscriptstyle\rm WA}$ is that it is not unexpectedly large
because no significant excess was found in the high-$q^2$ region by CLEO
\cite{cleo_WA}. In the context of our model of the high-$q^2$ tail,
this experimental result can actually be approximately translated into
an upper bound on $X$, which in turn is related to
$B_{\scriptscriptstyle\rm WA}$ or  $\mu_f$.  The CLEO bound is not
very strong, $X<0.07$ at 90\%CL, and still falls short of the expected
size for the WA contribution to the total rate $\delta
\Gamma_{\rm WA}/\Gamma_{\rm tot}\approx 2\%$ \cite{vub}.
We vary $X$ in the conservative range $0\le X\le 0.04$.   In terms
of model-independent parameters this corresponds to $0.008\GeV^3\le
B_{\scriptscriptstyle\rm WA}(1\GeV)\le0.020\GeV^3$.
In the future, it will be possible to measure the size of the WA expectation
values using, for instance, measurements of the $q^2$ spectrum close
to the endpoint and of its moments \cite{bumoments}, and by comparing
the semileptonic rates of $B^0$ and $B^+$.

In our default set-up we have assumed $X=0$, but while we expect it to
be positive and possibly small, there is no compelling reason for it
to vanish. The $X$ error in Table~1 should therefore not be treated as
gaussian.  It is worth observing that a non-zero value for $X$ would
bring the different experimental determinations of $|V_{ub}|$ closer
to each other.  For instance, $X=0.04$ would lead to $|V_{ub}|=
3.76,\, 4.21,\, 3.80 \times 10^{-3}$ in the {\bf A,B,C} cases,
respectively. Suggestive as this may sound, it is clear from our
analysis that an upper experimental cut on $q^2$ would improve the
theoretical precision in the inclusive determination of $|V_{ub}|$.
In the absence of a dedicated experimental analysis, we can subtract
Belle's results for the cases {\bf A} and {\bf B} \cite{belle1} and
extract a value of $|V_{ub}|$ for the combined cut $M_X\le1.7\GeV$,
$E_\ell>1.0 \GeV$, and $q^2<8\GeV^2$, i.e.\ with an {\it upper} cut
on $q^2$.  The cut on $q^2$ is relatively low and the predicted
fraction of events in this case is only 33\%, but the measured rate is
lower than expected: it corresponds to  $|V_{ub}|=3.18\times 10^{-3}$ 
with $\pm 4.5^{+1.7}_{-2.6}$\% theoretical error, 
dominated by parametric uncertainties.  We were
unable to compute the experimental uncertainty for the $q^2<8\GeV^2$
case, which will certainly be larger than those in Table~1.  Such a
low value of $|V_{ub}|$ may, however, indicate either an experimental
problem or an underestimate of theoretical errors in the high $q^2$
region, which is common to the three cases considered in Table~1.

\section{Conclusions}

We have calculated the triple differential distribution for inclusive
semileptonic decays without charm, $B\to X_u \ell\nu$, consistently
including all the known perturbative and non-perturbative effects,
through $O(\as^2\beta_0)$ and $O(1/m_b^3)$, respectively.  Our
theoretical framework is based on the OPE and incorporates a hard
Wilson cutoff $\mu\sim 1\GeV$.  This involved new perturbative
calculations discussed in Section 2.

Our approach has several new elements that we have listed in the
Introduction and explained in detail throughout the paper. We
recall the main ones: we parameterize the Fermi motion in terms of a
single light-cone function for each structure function and for any
value of $q^2$, and include consistently the subleading effects; this
is accomplished at the same level of model-dependence as for the
leading twist distributions; we implement for the first time the
complete BLM corrections to the triple differential rate; we present a
detailed discussion of the high-$q^2$ tail and of Weak Annihilation
effects.  Our approach is completely implemented in a C++ code that is
available from the authors.

We have extracted $|V_{ub}|$ from some of the latest and most precise
experimental data, providing a detailed estimate of the theoretical
uncertainty.  Our results, listed in Table~1, agree within theoretical
errors with those obtained with other methods used by HFAG
\cite{DGE,BLNP}.  We find that the dependence on the functional form
assumed for the distribution functions is rather weak.  However, the
critical role played by the high-$q^2$ tail becomes evident from our
analysis; it contributes in a significant way to the theoretical
uncertainty of the present inclusive determinations of $|V_{ub}|$. We
have modeled the high-$q^2$ region in two different ways complying
with the positivity of the differential spectra and have accounted for
the WA contributions. We find that non-vanishing WA effects tend to
suppress the value of $|V_{ub}|$ extracted from the data and, using
recent Belle results, we have argued that the low-$q^2$ sample of events
leads to a lower value of $|V_{ub}|$ that conflicts with that extracted
from the $q^2>8\GeV^2$ sample.

Since the high-$q^2$ tail presently leads to a sizeable uncertainty,
we encourage our experimental colleagues to pursue analyses with an
{\it upper} cut on $q^2$, and to perform an accurate separate
measurement of this domain.

\section*{Acknowledgements}
P.~Gambino is grateful to Einan Gardi for clarifying discussions.
N.~Uraltsev thanks I.~Bigi for discussions and collaboration on
related issues and the Department of Theoretical Physics of the
University of Turin for hospitality during part of the work.  
This work was supported in part by the NSF grant
PHY-0355098, by MIUR under contract 2004021808-009, and by a European
Community's Marie-Curie Research Training Network under contract
MRTN-CT-2006-035505 `Tools and Precision Calculations for Physics
Discoveries at Colliders'.

\section*{\huge Appendices}
\appendix
\section{Perturbative corrections}\label{app:real_with_cutoff}
In this Appendix we provide all necessary 
analytic expressions for the perturbative 
corrections to the three structure functions
 in the scheme with a hard cutoff $\mu$.
In the absence of cutoff the $O(\alpha_s)$ real gluon emission terms 
can be gleaned from \cite{dfn,Gambino:2006wk}:
\begin{align} \label{w1r}
\widetilde{R}_1^{(1)}(\hat{q}_0,\hat{q}^2) &= 
\frac{\ln \xi  
}{8\sqrt{\hat q_0^2-\hat q^2}}
\left[
   \frac{(\hat q_0+5) (\hat q_0-1)^3}{\hat q^2-\hat q_0^2}+\hat q_0^2-2
   \hat q_0-\hat q^2-14
\right] 
\notag\\& 
+\frac{(\hat q_0+5) \hat u}{4
   \left(\hat q_0^2-\hat q^2\right)}+\frac{5}{2}
- 2 \sqrt{\qzh^2 -\qqh}  \left( \frac{\ln \hat{u}}{\hat u} \right)_+ 
\notag\\& 
+ \left[\frac{7 (\hat q_0-1)}{2} + 4 \sqrt{\qzh^2 -\qqh} \, \ln (1-\hat q_0+\sqrt{\hat q_0^2-\hat q^2}) \right] \left(\frac{1}{\hat u} \right)_+,
\end{align}
\begin{align}
 \widetilde{R}_2^{(1)}(\hat{q}_0,\hat{q}^2) &=
-\frac{3 (\hat q_0+5) \,\hat u\,
   \hat q_0^2}{4 \left(\hat q_0^2-\hat q^2\right)^2}-\frac{6 \hat q_0^2-3
   \hat q^2\, \hat q_0+41 \hat q_0-\hat q^2-5}{4
   \left(\hat q_0^2-\hat q^2\right)} 
\notag\\& 
+ \frac{4 (\hat q_0-1)}{\sqrt{\hat q_0^2-\hat q^2}} 
\left( \frac{\ln \hat{u}}{\hat u} \right)_+
+\frac{\ln \xi
}{\sqrt{\hat q_0^2-\hat q^2}}
\Bigl[
\frac{32 \hat q_0^4+12 (\hat u-8) \hat q_0^3}{8 \left(\hat q_0^2-\hat q^2\right)^2} 
\notag\\& 
+ \frac{2 \left(\hat u^2-16
   \hat u+48\right) \hat q_0^2+2 \left(4 \hat u^2+7 \hat u-16\right)
   \hat q_0+\hat u \left(\hat u^2-7 \hat u+6\right)}{8 \left(\hat q_0^2-\hat q^2\right)^2}
\Bigr] 
\notag\\& 
- \left[ 7 + \frac{8 (\qzh -1)}{\sqrt{\qzh^2-\qqh}} 
\ln (1-\hat q_0+\sqrt{\hat q_0^2-\hat q^2}) \right] \left(\frac{1}{\hat u} \right)_+ ,
\label{w2r}
\\\notag\\
 \widetilde{R}_3^{(1)}(\hat{q}_0,\hat{q}^2) & =
-\frac{3
   \hat q_0+\hat q^2}{2 \left(\hat q_0^2-\hat q^2\right)}
+ \frac{2 (\hat q_0-1)}{\sqrt{\hat q_0^2-\hat q^2}} \left( \frac{\ln \hat{u}}{\hat u} \right)_+   
+\frac{\ln \xi
}{\sqrt{\hat q_0^2-\hat q^2}}
\left[
\frac{\hat q_0 (2 \hat q_0+\hat q^2-3)}{4
   \left(\hat q_0^2-\hat q^2\right)}
\right] 
\notag\\&
 - \left[ \frac{7}{2} + \frac{4 (\qzh -1)}{\sqrt{\qzh^2-\qqh}} 
\ln (1-\hat q_0+\sqrt{\hat q_0^2-\hat q^2}) \right] \left(\frac{1}{\hat u} \right)_+ ,
\label{w3r}
\end{align}
where $\hat u= 1-2\hat q_0+\hat q^2$ and 
\be
\xi= \frac{1-\hat q_0-\sqrt{\hat q_0^2-\hat q^2}}
{1-\hat q_0+\sqrt{\hat q_0^2-\hat q^2}}.
\ee
The plus distributions in Eqs.~(\ref{w1r}--\ref{w3r}) are defined by ($n\ge0$)
\begin{align} \label{plus_dist_log}
&\int_{a}^{\frac{1+\qqh}{2}} d \qzh \ G(\qzh,\qqh) 
\left[ \frac{\ln^n \left(1+\qqh-2\qzh \right)}{1+\qqh-2\qzh} \right]_+ =
 G \left( \frac{1+\qqh}{2},\qqh \right) \frac{\ln^{n+1} 
\left(1+\qqh-2a \right)}{2(n+1)}\notag\\  
& \qquad + \int_{a}^{\frac{1+\qqh}{2}} d \qzh 
\ \left[ G(\qzh,\qqh) - G \left( \frac{1+\qqh}{2},\qqh \right) \right]
\frac{\ln^n \left(1+\qqh-2\qzh \right)}{1+\qqh-2\qzh},
\end{align}
where  $G(\qzh,\qqh)$ is  a smooth function.

The NLO real emission contributions to the 
structure functions in the presence of  a Wilsonian cutoff, see 
\eqref{real}, are a new result and read as follows:
\begin{align}
 R_1^{cut,(1)}(\hat{q}_0,\hat{q}^2,\eta)& = 
\frac{(\hat q_0-1) \left(2
   \hat q_0^2+(\hat q^2+3) \hat q_0-5 \hat q^2-1\right)}{8
   \left(\hat q_0^2-\hat q^2\right)^{3/2} \eta }
-  \frac{\left(2
   \hat q_0^2+(\hat q^2-3) \hat q_0-2 \hat q^2+2\right) \eta }{2
   \left(\hat q_0^2-\hat q^2\right)^{3/2}} 
\notag\\
& +  \left[ \frac{2 (\hat q_0-1)^2}{\hat u_+}+\frac{(\hat q_0+5)
   (\hat q_0-1)^3}{ 8\,(\hat q_0^2-\hat q^2)}-\frac{\hat q_0^2}{8}+\frac{
   \hat q_0}{4}+\frac{\hat q^2}{8}-\frac14  
\right]
\frac{\ln \frac{1-\hat q_0+\sqrt{\hat q_0^2-\hat q^2}}{2\,\eta}}{\sqrt{\hat q_0^2-\hat q^2}}\notag\\
& +  \frac{5-\hat q_0}{8}+\frac{-\hat q_0^3-3 \hat q_0^2+9 \hat q_0-5}{8
   \left(\hat q^2-\hat q_0^2\right)}+\frac{7 (\hat q_0-1)}{4
   \hat u_+}
-\frac{(\hat q^2-1) \eta ^2}{4
   \left(\hat q_0^2-\hat q^2\right)^{3/2}}
\notag\\
&
 +  \frac{2 \hat q_0^3-(3 \hat q^2+1)
   \hat q_0^2+(4 \hat q^2-6) \hat q_0+\hat q^2+3}{16
   \left(\hat q_0^2-\hat q^2\right)^{3/2}}
\notag\\&
+  \frac{ \hat u\left( 2+\eta \right)}{16 \eta \sqrt{\hat q_0^2-\hat q^2}}
+\frac{(\hat q_0-1)^2-2 \eta ^2-8
   (\hat q_0-1) \eta }{4\sqrt{\hat q_0^2-\hat q^2}\, \hat u_+}
\end{align}
\begin{align}
 R_2^{cut,(1)}(\hat{q}_0,\hat{q}^2,\eta)& =
\frac{\ln \frac{1-\hat q_0+\sqrt{\hat q_0^2-\hat q^2}}{2\,\eta}}{\sqrt{\hat q_0^2-\hat q^2}}
\Bigl[\frac{4-4 \hat q_0}{\hat u_+}+
\frac{4-4   \hat q_0^2-2\hat q_0-\hat q^2 }{8} \notag\\
& +  \frac{-7 \hat q_0^4-8
   \hat q_0^3+56 \hat q_0^2-46 \hat q_0+5}{8
   \left(\hat q^2-\hat q_0^2\right)} 
- \frac{3 \left(\hat q_0^6+2 \hat q_0^5-12
   \hat q_0^4+14 \hat q_0^3-5 \hat q_0^2\right)}{8
   \left(\hat q^2-\hat q_0^2\right)^2}
 \Bigr]
\notag\\
&
-  \frac{3 \hat q_0+1}{8} +\frac{-6 \hat q_0^3-10 \hat q_0^2+41
   \hat q_0-5}{8 \left(\hat q^2-\hat q_0^2\right)}-\frac{7}{2 \hat u_+}-\frac{3
   \left(1-\qzh\right)^2 \left(\hat q_0^3+5 \hat q_0^2\right)}{8
   \left(\hat q^2-\hat q_0^2\right)^2}
\notag\\
& +  \frac{\left(2 \hat q_0^3+2 \hat q_0^2+\hat q^2 (3 \hat q^2-5) \hat q_0-3
   \hat q^4+\hat q^2\right) \eta ^2}{4 (\hat q_0-1)
   \left(\hat q_0^2-\hat q^2\right)^{5/2}}
-\frac{\hat q^2 (9 \hat q^2+1)}{8
   \left(\hat q_0^2-\hat q^2\right)^{5/2} \eta}
\notag\\
&
+  \frac{\left(4 \hat q_0^3+2
   (\hat q^2+2) \hat q_0^2+\hat q^2 (3 \hat q^2-13) \hat q_0-2 (\hat q^2-1)
   \hat q^2\right) \eta }{2 \left(\hat q_0^2-\hat q^2\right)^{5/2}}
\notag\\
& +  \frac{6
   \hat q_0^5-(3 \hat q^2+7) \hat q_0^4+(8-12 \hat q^2) \hat q_0^3+6 \left(2
   \hat q^4+\hat q^2-1\right) \hat q_0^2-2 \hat q^2 (6 \hat q^2-5)
   \hat q_0}{16
   \left(\hat q_0^2-\hat q^2\right)^{5/2}}
\notag\\
&
+  \frac{(\hat q^2-3) \hat q^2}{16
   \left(\hat q_0^2-\hat q^2\right)^{5/2}}
+\frac{\hat u\left(3\eta-2  \right)}{16 \eta\sqrt{\hat q_0^2-\hat q^2}}
+\frac{\frac{\eta ^2}{(\hat q_0-1)}+4 \eta
   +\frac{1-\hat q_0}{2}}{ \sqrt{\hat q_0^2-\hat q^2}\, \hat u_+}
\nonumber\\
&
+  \frac{-2 \hat q_0^5+(\hat q^2+1)
   \hat q_0^4+4 \hat q^2 \hat q_0^3-2 \left(2 \hat q^4+5 \hat q^2+1\right)
   \hat q_0^2+2 \hat q^2 (5 \hat q^2+6) \hat q_0}{8
   \left(\hat q_0^2-\hat q^2\right)^{5/2} \eta },
\end{align}
\begin{align}
 R_3^{cut,(1)}(\hat{q}_0,\hat{q}^2,\eta)&  =
\left[\frac{2-2
   \hat q_0}{\hat u_+}-\frac{\hat q_0 (2 \hat q_0+\hat q^2-3)}{4
   \left(\hat q_0^2-\hat q^2\right)}
 \right]
\frac{\ln \frac{1-\hat q_0+\sqrt{\hat q_0^2-\hat q^2}}{2\,\eta}}{\sqrt{\hat q_0^2-\hat q^2}}
\notag\\
& + \ \frac{\hat q_0^2+3 \hat q_0}{4
   \left(\hat q^2-\hat q_0^2\right)}-\frac{7}{4 \hat u_+} +\frac{1}{4} +
\frac{\hat q_0 \eta ^2}{2 (\hat q_0-1)
   \left(\hat q_0^2-\hat q^2\right)^{3/2}}+\frac{(2 \hat q_0+\hat q^2) \eta }{2
   \left(\hat q_0^2-\hat q^2\right)^{3/2}} \notag\\
& + \ \frac{\frac{\eta ^2}{2 (\hat q_0-1)}+ 2 \eta+\frac{1-\hat q_0}{4
   }}{\sqrt{\hat q_0^2-\hat q^2} \ \hat u_+} 
-\frac{(2 \hat q_0-1)   (\hat q_0-\hat q^2)}{8
   \left(\hat q_0^2-\hat q^2\right)^{3/2}}
-\frac{\hat q^2 \hat q_0+\hat q_0-2
   \hat q^2}{4 \left(\hat q_0^2-\hat q^2\right)^{3/2} \eta }.
\end{align}

We also report the soft-virtual structure functions
 in the absence of the cutoff 
\cite{dfn,Gambino:2006wk}. They are  consistent with the way we 
have performed the subtraction in the real emission contributions. At $O(\as)$
they are
\begin{eqnarray} 
\widetilde{V}_1^{(1)}(\hat{q}^2) &=& -\frac{1}{4} \left(1-\hat{q}^2\right) 
\Bigl[8 \ln ^2\left(1-\hat{q}^2\right)+2
  \left(\frac{1}{\hat{q}^2}-5\right) \ln \left(1-\hat{q}^2\right) \nonumber\\
  && \qquad +\ 4 \ \text{Li}_2\left(\hat{q}^2\right)+\frac{4 \pi ^2}{3}+5\Bigr] \label{w1v}\\
\widetilde{V}_2^{(1)}(\hat{q}^2) &=& 
-8 \ln ^2\left(1-\hat{q}^2\right)+10 \ln \left(1-\hat{q}^2\right) \nonumber
-  4 \text{Li}_2\left(\hat{q}^2\right)-\frac43 \pi^2-5
\label{w2v}\\
\widetilde{V}_3^{(1)}(\hat{q}^2) &=&\frac{2}{1-\qqh}
\,\widetilde{V}_1^{(1)}(\hat{q}^2) 
\label{w3v}
\end{eqnarray}
while at $O(\alpha_s^2 \beta_0)$ they are given by
\begin{eqnarray} 
\widetilde{V}_1^{(2)}(\hat{q}^2) &=&
  \frac{\left(1-\hat{q}^2\right)}4 
\left[
\left(\frac{1}{2 \qqh}-\frac{23}{6} -\ln \qqh\right) \ln^2(1-\qqh)
+\left(\frac{71 \qqh-19}{12 \qqh}+\frac{2 \pi ^2}{3}\right)
   \ln (1-\qqh)\right.\nonumber
\\
&&\quad \quad\quad\left.+\left(\frac{1}{2 \qqh}-\frac{19}{6}\right)
   \text{Li}_2(\qqh)-2 \text{Li}_3(1-\qqh)-\text{Li}_3(\qqh)+
\zeta(3)-\frac{79 \pi ^2}{72}-\frac{71}{24}\right] \nonumber\\
\widetilde{V}_2^{(2)}(\hat{q}^2) &=& 
-\left( \ln \qqh+\frac{23}{6}\right) \ln ^2(1-\qqh)
+\left(\frac{2 \pi
   ^2}{3}+\frac{71}{12}+\frac{1}{2\qqh}\right) \ln (1-\qqh)-\frac{19
   }{6}\text{Li}_2(\qqh)\nonumber
\\ &&-2 \text{Li}_3(1-\qqh)- \text{Li}_3(\qqh)+ \zeta
   (3)-\frac{79 \pi ^2}{72}-\frac{71}{24}\nonumber
\\
\widetilde{V}_3^{(2)}(\hat{q}^2) &=& \frac{2}{1-\qqh}
\,\widetilde{V}_1^{(2)}(\hat{q}^2)
\end{eqnarray}

\section{Structure functions in the local OPE and their   \\ $q_0$-moments}
 \label{app:q0moments}
In the adopted normalization, the power corrections to the structure functions
read
\begin{align}
 W_1^{pow}(\qzh,\qqh) =& \frac{\mug}{3m_b^2}\Bigl\{2 \delta _1 \left(2 \hat{q}^2-5 \hat{q}_0^2+7
   \hat{q}_0-4\right)-\delta _0\Bigr\} \nonumber\\
& +\frac{\mupi}{3m_b^2}\Bigl\{\delta _0-4 \delta _2 \left(\hat{q}_0-1\right)
   \left(\hat{q}_0^2-\hat{q}^2\right)+2 \delta _1 \left(\hat{q}_0 \left(5
   \hat{q}_0-3\right)-2 \hat{q}^2\right)\Bigr\} \nonumber\\
& +\frac{\rd}{9m_b^3}\Bigl\{-3 \delta _0+6 \delta _1
   \left(-\hat{q}^2+\left(\hat{q}_0-1\right) \hat{q}_0-2\right)\nonumber\\
& \quad + \ 4 \left(3
   \delta _2-2 \delta _3 \left(\hat{q}_0-1\right)\right)
   \left(\hat{q}_0-1\right) \left(\hat{q}_0^2-\hat{q}^2\right)\Bigr\}\nonumber\\
& +\frac{\rls}{3m_b^3}\Bigl\{-\delta _0+2 \delta _1
   \left(-\hat{q}^2+\left(\hat{q}_0-1\right) \hat{q}_0+2\right)+4 \delta _2
   \left(\hat{q}_0-1\right) \left(\hat{q}_0^2-\hat{q}^2\right)\Bigr\} 
\label{W1nonp}
\end{align}
\begin{align}
W_2^{pow}(\qzh,\qqh)= &\frac{2\mug}{3m_b^2} \Bigl\{2 \delta _1 \left(5 \hat{q}_0-2\right)-5 \delta _0\Bigr\}
 +\frac{2\mupi}{3m_b^2}\Bigl\{5 \delta _0-14 \delta _1 \hat{q}_0+4 \delta _2
   \left(\hat{q}_0^2-\hat{q}^2\right)\Bigr\}
\nonumber\\
& +\frac{2\rd}{9m_b^3}\Bigl\{3 \delta _0+6 \delta _1 \left(\hat{q}_0-4\right)+8
   \left(\hat{q}_0-1\right) \left(\delta _3
   \left(\hat{q}_0^2-\hat{q}^2\right)-3 \delta _2 \hat{q}_0\right)\Bigr\}
\nonumber\\
& +\frac{2\rls}{3m_b^3}\Bigl\{\delta _0+2 \delta _1 \left(\hat{q}_0-2\right)-2 \delta _2
   \left(\hat{q}^2+2 \left(\hat{q}_0-1\right) \hat{q}_0\right)\Bigr\}
\label{W2nonp}
\end{align}
\begin{align}
 W_3^{pow}(\qzh,\qqh)=& \frac{2\mug}{3m_b^2} \delta _1 \left(5 \hat{q}_0-6\right)
 +\frac{2\mupi}{3m_b^2}\Bigl\{2 \delta _2 \left(\hat{q}_0^2-\hat{q}^2\right)-5 \delta _1
   \hat{q}_0\Bigr\}-\frac{2\rls}{3m_b^3}\Bigl\{2 \delta _2 \left(\hat{q}_0-1\right)^2+\delta _1
   \hat{q}_0\Bigr\}
\nonumber\\
& +\frac{2\rd}{9m_b^3}\Bigl\{2 \left(\hat{q}_0-1\right) \left(2 \delta _3
   \left(\hat{q}_0^2-\hat{q}^2\right)-3 \delta _2 \hat{q}_0\right)-3 \delta _1
   \hat{q}_0\Bigr\}
\label{W3nonp}
\end{align}
We have used the short-hand notation
 $\delta_n = \delta^{(n)}(1+\qqh-2\qzh)$, where the $n$-th derivative of the
Dirac delta is taken wrt its argument.
We give explicit expressions for the zeroth, first and second 
$q_0$-moments, at fixed $q^2$, of the three form factors,
up to ${\cal O} (1/m_b^3)$ corrections in the
OPE. Separating the tree-level
and power corrections contributions, we define 
\begin{equation}
M_i^{(j),{\rm tree}} (\qqh) = \int d\qzh \  \qzh^j \ W_i^{\rm tree}( \qqh)\,
\delta(1+\qqh-2\qzh)= \frac12\left(\frac{1+\qqh}2\right)^j W_i^{\rm tree}( \qqh),
\end{equation}
and
\begin{equation}
M_i^{(j),pow} (\qqh) = \int d\qzh \  \qzh^j \ W_i^{pow}(\qzh, \qqh).
\end{equation}

The functions $I$'s defined in Sec.~\ref{sec:SF_derivation} correspond to 
central moments and are linear combinations of above $M_i^{(j),{\rm {\rm tree}}}$ or 
 $M_i^{(j),pow}$  moments:
\begin{eqnarray}
I_i^{(0)} &=& M_i^{(0)},\nonumber\\
I_i^{(1)} &=&  M_i^{(1)} -
\frac{1+\qqh}{2} \,  M_i^{(0)} \nonumber\\
I_i^{(2)} &=&  M_i^{(2)} -
\left( 1+\qqh \right)  \,  M_i^{(1)} +  
\left( \frac{1+\qqh}{2} \right)^2 \,  M_i^{(0)}. 
\end{eqnarray}

The explicit expressions for the zeroth moments are: 
\begin{equation}
M_1^{(0),{\rm tree}}=\frac{\left(1-\qqh\right)}{2}, \quad\quad
M_2^{(0),{\rm tree}}=2, \quad\quad M_3^{(0),{\rm tree}}= 1.
\end{equation}
\begin{eqnarray} 
M_1^{(0),pow}  &=& \frac{\left(1-5 \qqh\right)
   \mug}{6 m_b^2}+\frac{\left(\qqh+1\right) \mupi}{3
   m_b^2}+\frac{2 \qqh \rls}{3 m_b^3},\nonumber\\
M_2^{(0),pow}   &=& 0, \nonumber\\
M_3^{(0),pow}   &=& \frac{5 \mug}{6 m_b^2}-\frac{\mupi}{2
   m_b^2}-\frac{\rd}{6 m_b^3}-\frac{\rls}{2
   m_b^3}.
\end{eqnarray}
The explicit expressions for the first moments are:
\begin{eqnarray}
M_1^{(1),{\rm tree}}& =& \frac{1}{4} \left(1- \hat{q}^4\right), \quad\quad
M_2^{(1),{\rm tree}}  =   \left( \qqh+1\right), \quad\quad
M_3^{(1),{\rm tree}}  =  \frac{\left(\qqh+1\right)}2,\nonumber \\
M_1^{(1),pow}  &=& - \ \frac{\left(15
   \hat{q}^4-4 \qqh+5\right) \mug}{24 m_b^2} 
 + \frac{\left(3
   \hat{q}^4+8 \qqh+5\right) \mupi}{24 m_b^2}\nonumber\\
&& -  \frac{\left(5
   \hat{q}^4+7\right) \rd}{24 m_b^3}-\frac{\left(-15
   \hat{q}^4-5\right) \rls}{24 m_b^3},\nonumber\\
M_2^{(1),pow}   &=& \frac{\left(5
   \qqh+1\right) \mug}{6 m_b^2}-\frac{\left(\qqh+1\right)
   \mupi}{2 m_b^2} -  \frac{\left(\qqh+5\right) \rd}{6
   m_b^3}-\frac{\left(\qqh+1\right) \rls}{2 m_b^3},\nonumber\\
M_3^{(1),pow}  &=& - \ \frac{\left(1-5 \qqh\right)
   \mug}{6 m_b^2}-\frac{\left(\qqh+1\right)\mupi}{3 m_b^2}
   -\frac{2 \qqh \rls}{3 m_b^3}.\nonumber
\end{eqnarray}
The explicit expressions for the 
second moments are:
\begin{eqnarray}
M_1^{(2),{\rm tree}} = \frac{\left(1-\qqh\right)
   \left(\qqh+1\right)^2}{8},\quad
M_2^{(2),{\rm tree}} = \frac{\left(\qqh+1\right)^2}2,\quad
M_3^{(2),{\rm tree}}=\frac{\left(\qqh+1\right)^2}4,
\end{eqnarray}
\begin{eqnarray}
M_1^{(2),pow}  &=& -  \left(5
   \hat{q}^6+\hat{q}^4-\qqh+3\right)  \frac{\mug}{12 m_b^2}
 +  \left(2   \hat{q}^4+\qqh+1\right) \frac{\mupi}{6m_b^2} 
\nonumber\\ &&
   -\frac{\left(\hat{q}^6+\qqh+1\right) \rd}{3
   m_b^3}
+ \ \frac{\left(3 \hat{q}^6+\hat{q}^4+\qqh+1\right) \rls}{6
   m_b^3},\nonumber\\
M_2^{(2),pow}  &=& 
   \left(5 \hat{q}^4+6 \qqh+1\right) \frac{\mug}{6m_b^2}
-  \left(\hat{q}^4+4 \qqh+1\right) \frac{\mupi}{3m_b^2}
-  \frac{\left(4
   \qqh+2\right) \rd}{3 m_b^3}
-\frac{\left(2 \hat{q}^4+4
   \qqh+1\right) \rls}{3 m_b^3},\nonumber\\
M_3^{(2),pow}   &=& \frac{\left(5
   \hat{q}^4+2 \qqh-3\right) \mug}{8 m_b^2}-\frac{\left(3
   \hat{q}^4+14 \qqh+3\right) \mupi}{24 m_b^2}\nonumber\\
&&+ \ \frac{\left(5
   \hat{q}^4-2 \qqh+1\right) \rd}{24
   m_b^3}+\frac{\left(-15 \hat{q}^4-6 \qqh+1\right) \rls}{24 m_b^3}. 
\nonumber
\end{eqnarray}

\section{Local OPE results with arbitrary cuts}
In this Appendix we report the local OPE results 
for the rate of $B\to X_u \ell\nu$ subject to standard
cuts on $M_X$, $E_\ell$, and $q^2$. The expressions contain 
only the non-perturbative power corrections. We adopt the following 
notation
\be 
\xi=\frac{2E_{\ell,cut}}{m_b} \quad\quad 
\tau=max\left(\frac{q^2_{cut}}{m_b^2},\frac{M_B \bar\Lambda-
M_{X,cut}^2}{m_b\bar\Lambda}\right),
\ee
where $E_{\ell,cut}$ and $ q^2_{cut}$ are lower cuts on $E_\ell,q^2$, and 
$M_{X,cut}^2$ is an upper cut on the invariant mass of the hadronic system.
Only $\tau$, the {\it effective lower cut on $q^2$}, is relevant for 
power corrections: the upper cut on the hadronic mass must 
be lower than $\sqrt{\bar \Lambda \, M_B}\approx 1.8\GeV$ to play a role in 
the following expressions. There is also a relation between $\xi$ and $\tau$:
both cuts are relevant only when $\tau<\xi\le 1$, otherwise only $\tau$ is 
relevant. The total width subject to the above cut is given in the two regions
 by 
\begin{eqnarray} \label{rateCut1}
\frac{\Gamma_{cut}}{\Gamma_0} ( \xi > \tau)& =& 
 (\xi -1) \left(\xi ^3-4 \tau  \xi ^2-\xi ^2+6 \tau ^2
   \xi +2 \tau  \xi -\xi -4 \tau ^3+2 \tau -1\right) \nonumber\\
&& -  \frac{\mu _{\pi }^2}{6 m_b^2}\Bigl[-5 \xi ^4+6 (3 \xi -4) \tau ^2 \xi
-4 (4 \xi   -3) \tau ^3
 + \ 6 \left(3 \xi ^2-1\right) \tau +3\Bigr]\nonumber \\
&& +  \frac{\mu _G^2}{6 m_b^2}\Bigl[-5 \xi ^4-8 \xi ^3+6 (5 \xi +2) \tau ^2
\xi +(20-40 \xi ) \tau ^3  + \ 6 \left(\xi ^2-1\right) \tau
   -9\Bigr] \nonumber\\
&& +  \frac{\rho
   _{LS}^3}{6 m_b^3}\left[\xi ^4-18 \tau ^2 \xi ^2+4 (8 \xi -3) \tau
   ^3+\left(6-18 \xi ^2\right) \tau +9\right]  \nonumber\\
&& -  \frac{\rho
   _D^3}{6 m_b^3}\Bigl[-\xi ^4-16 \xi ^3+6 \tau ^2 \xi ^2+30 \tau
  \xi ^2+24 \xi ^2+48 \xi +4 \tau ^3 +26 \tau -77\nonumber \\
&& \quad\quad\quad \,-  48 \ln \frac{\muwa^2}{m_b^2} 
   +48 \ln (1-\xi )+48 \ln (1-\tau )\Bigr]
 +32\pi^2 \frac{B_{\scriptscriptstyle\rm WA}(\muwa)}{m_b^3}, 
\end{eqnarray}
and
\bea \label{rateCut2}
\frac{\Gamma_{cut}}{\Gamma_0} (\xi \le \tau)
&=&  -(\tau -1)^3 (\tau +1) +\frac{\mu _{\pi }^2}{2 m_b^2}
(\tau -1)^3 (\tau +1) 
-  \frac{\mu_G^2}{2 m_b^2}\left(5 \tau ^4-10 \tau ^3+2 \tau +3\right)
\nonumber \\
&&
+ \frac{\rho_{LS}^3}{2 m_b^3} \left(5 \tau ^4-10 \tau ^3+2 \tau +3\right)
+32\pi^2 \frac{B_{\scriptscriptstyle\rm WA}(\muwa)}{m_b^3} 
\nonumber \\
&& -  \frac{\rho_D^3}{6 m_b^3}\Bigl[5 \tau ^4+22 \tau ^3+24 \tau ^2+74
\tau -48 \ln \frac{\muwa^2}{m_b^2}  + \ 96 \ln (1-\tau )-77\Bigr]. 
\eea

\end{document}